\documentclass[aps,prb,twocolumn,groupedaddress,showpacs,10pt]{revtex4-1}

\usepackage{graphicx}
\usepackage{lmodern}
\usepackage{amsmath}

\newcommand{\comments}[1]{}
\begin{document}


\title{Probing the charge of a quantum dot with a nanomechanical resonator}

\author{H.B. Meerwaldt}
\email{H.B.Meerwaldt@tudelft.nl}

\author{G. Labadze}
\author{B.H. Schneider}
\author{A. Taspinar}
\author{Ya.M. Blanter}
\author{H.S.J. van der Zant}
\author{G.A. Steele}
\email{G.A.Steele@tudelft.nl}

\affiliation{Kavli Institute of Nanoscience, Delft University of Technology, Lorentzweg 1, 2628CJ Delft, The Netherlands}


\author{}
\affiliation{}


\date{\today}

\begin{abstract}
We have used the mechanical motion of a carbon nanotube (CNT) as a probe of the average charge on a quantum dot. Variations of the resonance frequency and the quality factor are determined by the change in average charge on the quantum dot during a mechanical oscillation. The average charge, in turn, is influenced by the gate voltage, the bias voltage, and the tunnel rates of the barriers to the leads. At bias voltages that exceed the broadening due to tunnel coupling, the resonance frequency and quality factor show a double dip as a function of gate voltage. We find that increasing the current flowing through the CNT at the Coulomb peak does not increase the damping, but in fact decreases damping. Using a model with energy-dependent tunnel rates, we obtain quantitative agreement between the experimental observations and the model. We theoretically compare different contributions to the single-electron induced nonlinearity, and show that only one term is significant for both the Duffing parameter and the mode coupling parameter. We also present additional measurements which support the model we develop: Tuning the tunnel barriers of the quantum dot to the leads gives a 200-fold decrease of the quality factor. Single-electron tunneling through an excited state of the CNT quantum dot also changes the average charge on the quantum dot, bringing about a decrease in the resonance frequency. In the Fabry-P\'{e}rot regime, the absence of charge quantization results in a spring behaviour without resonance frequency dips, which could be used, for example, to probe the transition from quantized to continuous charge with a nanomechanical resonator. \end{abstract}


\maketitle


\section{Introduction}

Nanomechanical systems\cite{Craighead2000,Ekinci2005} are studied intensively for both their potential applications such as mass sensing\cite{Lavrik2003,Ekinci2004, Lassagne2008} and for insights into the quantum mechanical ground state of a macroscopic object\cite{Schwab2005, O'Connell2010, Teufel2011, Chan2011,Poot2012}. Because of their small size, nanomechanical resonators are strongly influenced by electrostatic forces from single-electron charge effects, which allows single-electron transistors to be used as sensitive detectors of the deflection of a nanomechanical beam \cite{Knobel2003,LaHaye2004}, demonstrating clear backaction from the single-electron forces \cite{Naik2006}.
Coupling of these forces to mechanical resonators can also be exploited in mechanical single-electron shuttle devices \cite{Koenig2008} to shuttle electrons one-by-one through the nanomechanical resonator.

A carbon nanotube (CNT) is a stiff, bottom-up nanomechanical resonator with a large aspect ratio\cite{Sazonova2004c}.
Dissipation in an ultraclean CNT at cryogenic temperatures is low, which results in a high quality factor\cite{Huettel2009} and allows investigation into other sources of damping, such as nonlinear\cite{Eichler2011} and magnetic damping\cite{Schmid2012}. At cryogenic temperatures, a quantum dot is formed, embedded in the CNT\cite{Tans1997,Bockrath1997}, which makes single-electron charge effects couple strongly to the mechanical motion through the bending mode.\cite{Lassagne2009,Steele2009}. For a CNT quantum dot, the effects of damping, spring stiffening and softening, and nonlinearity, are completely dominated by single-electron charging effects\cite{Meerwaldt2012}.

The interplay between single-electron tunneling and mechanical motion has been the topic of many theoretical investigations. The motion of nanomechanical resonators is found to have an influence on the electron transport through the single-electron transistor \cite{Doiron2006,Chtchelkatchev2004} affecting current \cite{Armour2004,Labadze2011} and current noise \cite{Armour2004a,Clerk2005,Flindt2005,Brueggemann2012}. Conversely, transport through the single-electron transistor by tunneling of single electrons causes backaction\cite{Mozyrsky2004,Clerk2005,Pistolesi2007,Pistolesi2008} on the mechanical motion in the form of frequency shifts \cite{Rodrigues2005, Nocera2011, Meerwaldt2012} and damping \cite{Rodrigues2005,Rodrigues2007, Bennett2010,Labadze2011}. Under certain conditions, single-electron tunneling can lead to a negative charging energy \cite{Ojanen2009} or negative damping, causing instabilities, where a distinction should be made between the low-frequency limit\cite{Usmani2007} ($\omega_0 \ll \Gamma$)  and the high-frequency limit\cite{ElBoubsi2008} ($\omega_0 \gg \Gamma$), relating the mechanical resonance frequency, $\omega_0$, to the single-electron tunnel rate, $\Gamma$. Besides causing backaction on the mechanical motion, single-electron tunneling is proposed to be used to parametrically drive the nanomechanical resonator \cite{Midtvedt2011}. In the Coulomb blockade regime, CNT quantum dot resonators in particular are found to have a large electron-vibron coupling \cite{Mariani2009}, and nonlinear restoring forces are found to be completely dominated by single-electron tunneling effects \cite{Nocera2011,Meerwaldt2012}. Further theoretical studies have been performed on the subject of single-electron shuttles \cite{Gorelik1998, Armour2002,Pistolesi2005,Huldt2007}, and on the coupling between a single-electron transistor and a nanomechanical resonator in the quantum regime \cite{Armour2002a, Armour2008,Blencowe2008}.

In this article, we present measurements of single-electron effects in CNT resonators in the Coulomb blockade regime and we demonstrate that our experimental observations agree quantitatively with the theoretical model we develop. This agreement allows the CNT nanomechanical resonator to be used as a probe for the average charge residing on the CNT quantum dot. Furthermore, we examine the implications of the established model through additional experiments. The layout of the paper is as follows. In section \ref{sec:fabrication}, the fabrication of the ultraclean suspended CNT is described, followed by the measurement setup. Section \ref{sec:characterization} contains the characterization of the CNT device in electrical terms, and the influence of single-electron tunneling on the mechanical resonance frequency, followed by the novel experimental observation of a double frequency dip feature. In section \ref{sec:SEspring}, we develop a model for the dynamics of the mechanical resonator in the presence of Coulomb blockade, which explains the presence of such a double frequency dip, and make a quantitative comparison to the measured data. In section \ref{sec:SEdamping}, we extend this model to include the effect of Coulomb blockade on mechanical damping and, again, compare the model to experimental data. In section \ref{sec:SEnonlinearity}, we expand the model further with the description of single-electron induced nonlinearity, resulting in an expression for the Duffing parameter and the mode coupling parameter. In section \ref{sec:additional}, we explore the coupling of the average charge to the mechanical resonator by varying the tunnel rates of the quantum dot, by studying the effects of excited states of the quantum dot, and by studying the mechanical resonator in the Fabry-P\'{e}rot regime, in which Coulomb Blockade no longer plays a role. All of these additional measurements can be understood qualitatively in the context of the model we present. 

\section{Fabrication and measurement setup}
\label{sec:fabrication}
Ultraclean suspended carbon nanotube devices are fabricated as follows\cite{Steele2009a}. The fabrication begins with a degenerately doped silicon wafer with a 285 nm thermal oxide. In the first step, the contacts are patterned by evaporating 5 nm of tungsten and 25 nm of platinum on a patterned double layer of poly(methyl methacrylate) (PMMA), and performing lift-off. In the second step, a three-layer etch mask, consisting of photoresist, tungsten, and PMMA, is used to etch the trenches between the contacts. The trenches are first patterned onto the PMMA by electron beam lithography. This pattern is transferred onto the tungsten by parallel plate reactive ion etching using a mixture of SF$_6$ and helium, where the PMMA acts as an etch mask. After this, the photoresist is etched by an oxygen plasma, during which the tungsten acts as an etch mask. In the third step, trenches between the contacts are etched into the silicon oxide, by the same mixture of SF$_6$ and helium, during which the photoresist acts as an etch mask. To improve wire-bonding, a layer of 10 nm of chromium and 80 nm of platinum is evaporated onto the bondpads, followed by a sputtered layer of 20 nm of silicon. In the final step, catalyst islands\cite{Kong1998} are deposited onto holes patterned in a double layer of PMMA. The sample is now placed in a CVD oven, where CNTs grow out of the catalyst particles in a mixture of hydrogen and methane at a temperature of 900 $^\circ$C. As the CNTs grow in a random direction, approximately one third of the patterned trenches has a CNT across them, touching both the source and the drain, thus forming a device. Room temperature measurements of current as a function of gate voltage are performed for each trench, showing semiconductor behaviour for potential devices.

Figure \ref{fig1}a shows a schematic diagram of the setup used to measure mechanical resonances in the suspended carbon nanotube devices\cite{Huettel2009}. The device is mounted at the mixing chamber of a $^3$He/$^4$He dilution refrigerator with a temperature of 20 mK. Filtered twisted-pair cabling is used to connect to the source, drain, and gate of the device, allowing d.c. voltages to be applied to the source and gate. The current flowing through the device is measured at the drain. The CNT is driven into motion by an a.c. voltage difference between the gate electrode and the CNT. This high-frequency signal needed to drive the CNT is supplied by an RF source through a coaxial cable. At a separation of $\sim$1 cm from the device, the shielding of the coaxial cable is removed to form the antenna. We expect that the electrostatic coupling between the antenna and the nanotube segment itself is much too small to actuate the CNT, which was confirmed by a lack of response of the CNT quantum dot to a d.c. voltage applied to the coax. Instead, the coaxial cable is capacitively coupled to the d.c. wires leading to the source, drain, and gate. Because of a difference in crosstalk capacitance from the coaxial cable to the source, drain, and gate, a.c. voltages are generated asymmetrically on the three. The a.c. voltage difference arising between the gate and the CNT then actuates the CNT into motion.

\section{Electrical and mechanical characteristics}
\label{sec:characterization}
\begin{figure}[ht]
\centering
	\includegraphics{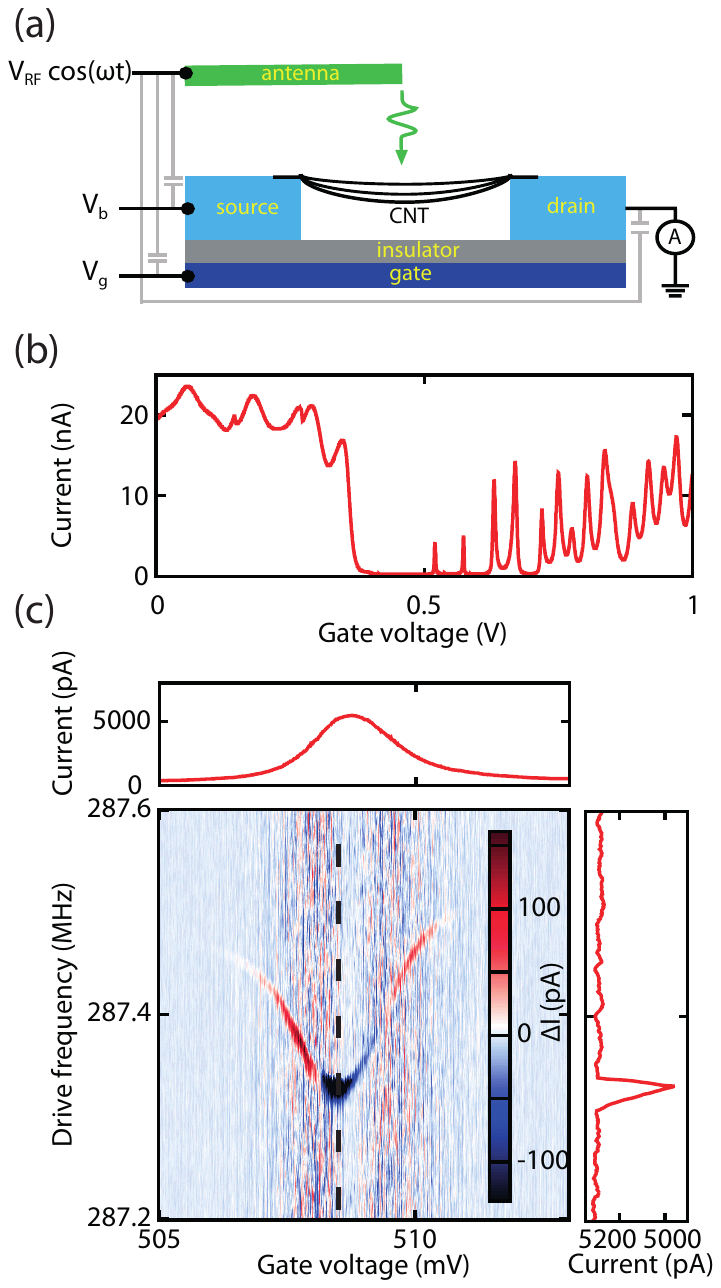}
	\caption{(a) Schematic drawing of the chip geometry, antenna, and measurement electronics. The CNT acts as a 
doubly-clamped beam resonator with a displacement $x$ and is driven because of an asymmetrical capacitive coupling of the radio frequency coaxial cable to the source, drain, and gate.
(b) Current versus gate voltage at $V_b = 0.3$ mV showing, for increasing gate voltage, Fabry-P\'{e}rot 
oscillations, then a small bandgap, and finally Coulomb oscillations with the increasing tunnel coupling to the leads opening up the quantum dot as the gate voltage is increased further.
(c) Top panel: Current as a function of gate voltage, showing a Coulomb peak. Middle panel: Change in current as a function of drive frequency and gate voltage, showing a dip in resonance frequency across the Coulomb peak. Right panel: Current as a function of drive frequency at a gate voltage, $V_g=508.5$ mV, as denoted by the black dashed line in the middle panel, showing the mechanical resonance as a decrease in current around $f=287.33$ MHz. 
}
\label{fig1}
\end{figure}

Figure \ref{fig1}b shows the current through a small-bandgap CNT with a suspended length of 600 nm as a function of gate voltage. For gate voltages below $V_g=0.4$ V, the device is doped with holes and weak scattering at the metal-CNT interface at the edge of the trench results in conductance that is modulated by Fabry-P\'{e}rot interferences\cite{Liang2001} of the hole wavefunction. For $0.4 < V_g < 0.5$, the Fermi level lies in the bandgap and the current is suppressed. From the distance in gate voltage from electron and hole conduction, together with the coupling factor $\alpha=C_g/C_{tot}=0.38$ determined from the Coulomb diamonds, we estimate the bandgap to be $E_{gap} = 58$ meV. Above $V_g=0.5$ V, Coulomb oscillations are visible. Now, electrons tunnel onto the CNT through tunnel barriers, which arise from the p-n junctions between the segments of the CNT near the W/Pt metal near the edge of the trench and the CNT. The small capacitance between the quantum dot and the three terminals gives rise to a large charging energy of $E_C = 9.6$ meV. For $V_g>0.7$ V, the Coulomb peaks are increasingly more smeared out. For larger electron doping, the p-n junctions become narrower and the tunnel barriers to the quantum dot become more transparent\cite{Steele2009a}.

We measure the mechanical bending mode resonances of the CNT by actuating it into motion and measuring the d.c. current. When the drive frequency matches the resonance frequency, the CNT resonates, causing the capacitance between the CNT and the gate to oscillate with a large amplitude. This oscillating capacitance effectively induces an oscillating gate voltage. The nonlinearity of the Coulomb peak allows the oscillating effective gate voltage from the motion to be rectified into a d.c. current\cite{Huettel2009}. 

As reported earlier, the mechanical resonance frequency of the CNT is strongly influenced by single-electron tunneling\cite{Steele2009}. The middle panel of Fig. \ref{fig1}c shows the change in current due to the mechanical motion, $\Delta I$, as a function of the drive frequency and gate voltage on a Coulomb peak at $V_b = 0.2$ mV. On a Coulomb peak, single-electron tunneling leads to a dip in the resonance frequency. In the top panel of Fig. \ref{fig1}c, a linecut is shown of current versus gate voltage at a drive frequency chosen to be far from the resonance frequency. In the right panel of Fig. \ref{fig1}c, a linecut is shown of the current versus the drive frequency at one gate voltage, denoted by the black dashed line in the middle panel, showing the mechanical resonance as a change in the d.c. current. 

In order to qualitatively understand these dips in frequency, three essential elements are required. The first is that the motion of the CNT changes the charge on the quantum dot, which in turn changes the electrostatic force on the CNT. Thus, the CNT experiences a displacement-dependent electrostatic force. Because the electron tunnel rates $\Gamma$ are much faster than the mechanical frequency $f_0$ ($\Gamma \sim 450$ GHz, $f_0 \sim 300$ MHz), the mechanical motion sees a displacement-dependent force from the charge averaged over many tunnel events. This displacement-dependent force induces a reduction of the spring constant, which changes the mechanical frequency. 

The sign of the frequency shifts can be understood by realizing that electrostatic forces do not act as a restoring force, but instead as an anti-restoring force: if the CNT is pulled away from its equilibrium position towards the gate, for example, the electrostatic force will increase and tend to pull the CNT with more force towards the gate. Thus, electrostatic forces generally result in a decrease of the net spring constant.

Finally, to understand the gate voltage dependence, it is illustrative to examine the differential capacitance of the quantum dot to the gate. Due to the quantized charge on the island of the dot, the quantum dot shows zero differential capacitance when it is in the Coulomb valleys: the charge is fixed independent of gate voltage. At the position of the Coulomb peaks, the charge on the island undergoes a discrete step of one electron. In the absence of tunnel or temperature broadening of this transition, this step would be infinitely sharp, and the differential capacitance would diverge. The electrostatic spring effects discussed in the previous paragraph are determined by the differential capacitances for small amplitudes of motion, and the diverging differential capacitance of the quantum dot at the positions of the Coulomb peaks leads to a diverging softening renormalization of the net mechanical spring constant, resulting in the dips in mechanical frequency observed in the data.

\begin{figure}[ht]
\centering
	\includegraphics{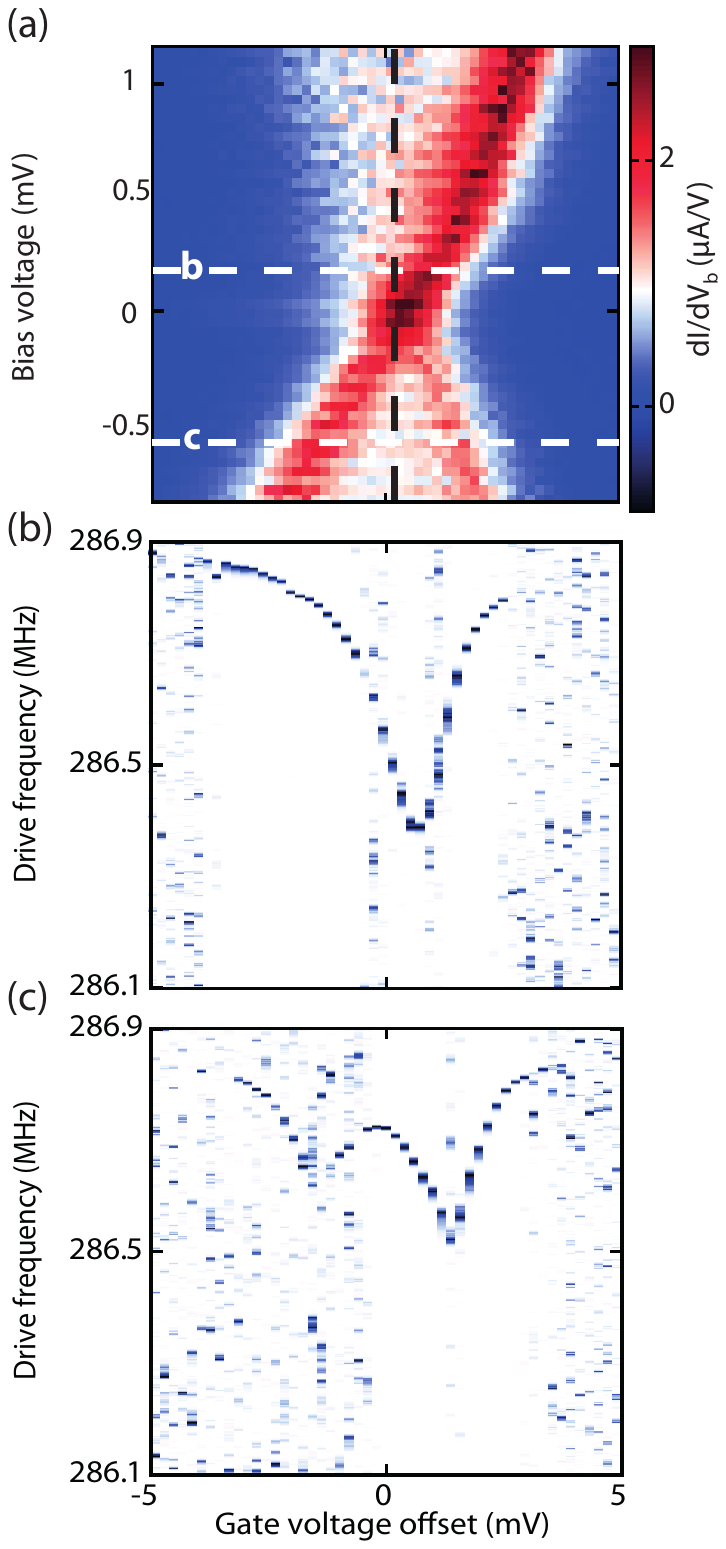}
	\caption{(a) Stability diagram: $dI/dV_b$ as a function of bias and gate voltage for the charge transition from 1 to 2 electrons. The linecuts, at which Figs. \ref{fig2new}b and c are taken, are indicated by the white dashed lines. The black dashed line denotes the gate voltage at which the middle panel of Fig. \ref{fig2}c is taken. (b, c) Measured normalized current, $|(I-I_{0})|/|I-I_0|_{max}$, as a function of drive frequency and gate voltage offset at (b) $V_b = 0.17$ mV, showing a decrease in resonance frequency around $V_{g,offset}=1$ mV, and at (c) $V_b = -0.55$ mV, showing a double dip in resonance frequency.
}
\label{fig2new}
\end{figure}

In this article, we explore in detail how single-electron tunneling affects the mechanical motion of the CNT and develop a quantitative model to describe our results. In contrast to earlier work\cite{Steele2009}, here we map out the behaviour of the mechanical resonator for bias and gate voltages covering the single-electron tunneling region between two charge states. The stability diagram in Fig. \ref{fig2new}a shows the differential conductance, $dI/dV_b$, as a function of gate voltage and bias voltage for the transition from 1 to 2 electrons. Coulomb blockade is visible in blue, whereas single-electron tunneling takes place in the red and white regions.

At a low bias voltage, Fig. \ref{fig2new}b shows, in blue, the experimentally obtained normalized current, $|(I-I_{0})|/|I-I_0|_{max}$, as a function of gate voltage and drive frequency, where $I_0$ is the current off mechanical resonance. The bias voltage of $0.17$ mV for this figure is denoted by the upper white dashed line in Fig. \ref{fig2new}a. The mechanical resonance is visible as an increase in the normalized current. Around $V_{g,offset}=1$ mV, where $V_{g,offset}=V_g - 0.565$ V, the mechanical resonance frequency shows a dip, as was demonstrated in previous measurements.

When the bias voltage is increased, the resonance frequency of the CNT exhibits substantially different behaviour. Figure \ref{fig2new}c shows the measured normalized current as a function of gate voltage and drive frequency as a color plot at a bias voltage of $-0.55$ mV, corresponding to the lower white dashed line in Fig. \ref{fig2new}a. Instead of one dip, two dips in resonance frequency are visible. The presence of such a peculiar double dip feature in the mechanical resonance frequency forms the motivation for the work in the following sections, in which we establish a quantitative model for the mechanical resonance frequency and quality factor in the presence of Coulomb blockade, and further explore the coupling of the mechanical resonator dynamics to the average charge of the quantum dot. 

\section{Modeling single-electron spring effects}
\label{sec:SEspring}
This section describes a model developed for the coupling of the mechanical resonator to the Coulomb blockaded quantum dot, and the quantitative fitting of the experimentally observed frequency dips. In subsection \ref{subs:modelavcharge}, a model for the average charge on the CNT is described. In subsection \ref{subs:modelspring}, we derive a model for how the average charge leads to a displacement-dependent force that causes the softening of the CNT spring. In subsection \ref{subs:fitspring}, fits of the experimentally obtained resonance frequency and current are performed using the established model.

\subsection{Model for the average charge on a quantum dot}
\label{subs:modelavcharge}
The average charge on the CNT at a charge transition can be determined as follows. Single-electron tunneling onto or off the CNT is a stochastic process, where the amount of time an electron spends on the source or the CNT is determined, respectively, by the tunneling rates $\Gamma^+$ and $\Gamma^-$. The average occupation, $\langle N\rangle$, of charges on the CNT for the charge transition from $N_0$ to $N_0+1$ electrons is given by:
\begin{equation}
\langle N\rangle = N_0+\frac{\Gamma^+}{\Gamma_{tot}},
\label{eq:averagecharge}
\end{equation}
where $\Gamma_{tot}=\Gamma^++\Gamma^-$. 

The tunneling rates, determining the average occupation, are modeled as follows. Traditionally, tunneling onto and off a quantum dot is described\cite{Beenakker1991, Thijssen2008} using energy-independent tunnel rates. The average occupation and the current are determined by using energy-independent tunnel rates and the overlap of the density of available states of the quantum dot and the leads. Such an approximation of energy-independent tunnel rates is valid when the tunnel barriers are sufficiently high, so that a change in either bias or gate voltage does not cause a significant change in the barrier height or width.

\begin{figure}[ht]
\centering
	\includegraphics{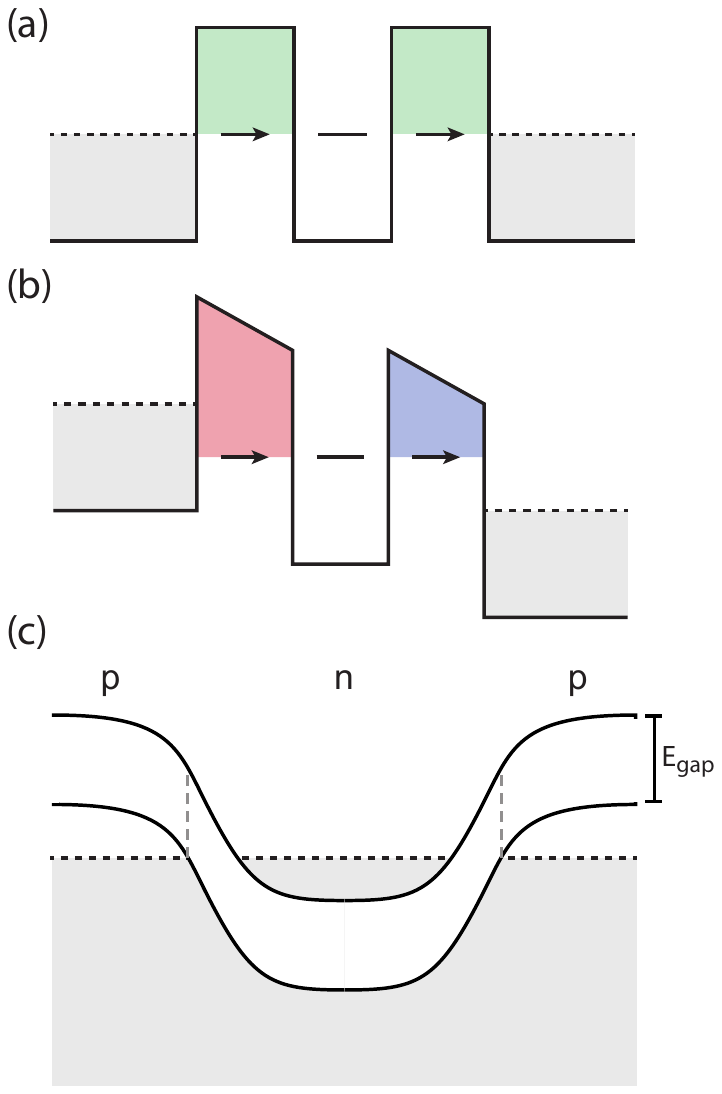}
	\caption{(a) Energy diagram at zero bias voltage, with the p-n junctions represented by square tunnel barriers between the CNT and the leads. (b) Energy diagram at finite bias voltage, showing an increase of the height of the tunnel barrier to the source relative to the energy of the tunneling electron, and a decrease in the height of the barrier to the drain. (c) Diagram of the CNT band structure with two p-n junctions induced through a positive gate voltage. }
	\label{fig3new}
\end{figure}

In this work instead, we use energy-dependent tunnel rates, assuming that the tunnel barriers are not high. This assumption is supported by the strong dependence of the tunnel barriers on the gate voltage (cf Fig. \ref{fig1}b). The influence of bias and gate voltage on the tunnel rates is demonstrated in Figs. \ref{fig3new}a and b. At zero bias voltage, Fig. \ref{fig3new}a shows the p-n junctions as tunnel barriers between the CNT and the leads. When the bias voltage is increased, Fig. \ref{fig3new}b illustrates how the higher electrostatic potential of the source increases the tunnel barrier at the source, leading to a lower tunnel rate, whereas the tunnel barrier at the drain is decreased, resulting in a higher tunnel rate. In our device, the tunnel barriers are not formed by steps in the potential but instead by p-n junctions in the CNT. Figure \ref{fig3new}c shows how tunnel barriers arise from p-n junctions formed in the CNT at the metal-CNT interface, as the chemical potential of the CNT is decreased by a positive gate voltage. The height of the tunnel barriers is determined by the bandgap of the CNT, $E_{gap}$.

To calculate the tunnel rates, the density of available states of the leads and the quantum dot are used, taking the energy dependence of the tunnel rates into account. The density of available states of the level in the CNT quantum dot, $D(\mu)$, caused by tunnel coupling to the leads, is described with a Lorentzian lineshape.
\begin{equation}
D(\mu)=\frac{1}{2\pi}\frac{\hbar\Gamma_{broad}}{(\mu-\mu_{CNT})^2+(\hbar\Gamma_{broad}/2)^2}.
\label{eq:broadening}
\end{equation}
Here, $\mu_{CNT}$ is the chemical potential of the CNT. The broadening $\Gamma_{broad}$ gives the full width at half maximum. At $T = 20$ mK, we are in the regime where $\hbar\Gamma_{broad}\gg k_B T$. The density of available states of the left and right lead, $f_{L,R}(E)$, is now modeled by a step function:
\begin{align}
   f_{L,R}(E) = \left\{
     \begin{array}{lr}
       1 : E < \mu_{L,R} \\
       0 : E > \mu_{L,R},
     \end{array}
   \right.
\end{align} 
where $E$ is the energy of the electron, and $\mu_{L,R}$ is the chemical potential of the left (right) lead. The tunnel rates are determined by the overlap of the density of available states of the CNT and the leads, and by their relative chemical potential:
\begin{equation}
\Gamma^{\pm}_{L,R}=a_{L,R}e^{b_{L,R}\Delta \mu_{L,R}}\left(\frac{1}{2}+\frac{1}{\pi}\arctan\left(\frac{\mp2\Delta \mu_{L,R}}{\hbar \Gamma_{broad}}\right)\right).
\label{eq:gammas}
\end{equation}
Here, the energy-dependent tunneling is reflected by the exponential factor, which is shown theoretically\cite{Korotkov1991} to occur for tunnel barriers brought about by an electrostatic potential, where the barrier height is low and the barrier width is large. The parameters $a_{L,R}$ depend on the height and width of the tunnel barriers. The parameters $b_{L,R}$ describe the triangular profile that the tunnel barriers have, which is there at zero bias voltage (cf Fig. \ref{fig3new}c) and is additionally changed by altering the bias and gate voltage (cf Fig. \ref{fig3new}b). The difference in chemical potential between the CNT and the left or right lead is denoted by $\Delta \mu_{L,R}=\mu_{CNT}-\mu_{L,R}$. The last factor in Eq. \ref{eq:gammas} arises from the broadening due to tunnel coupling and is determined from the overlap of the density of available states of the left or right electrode with that of the CNT. 

\begin{figure*}[ht]
\centering
	\includegraphics{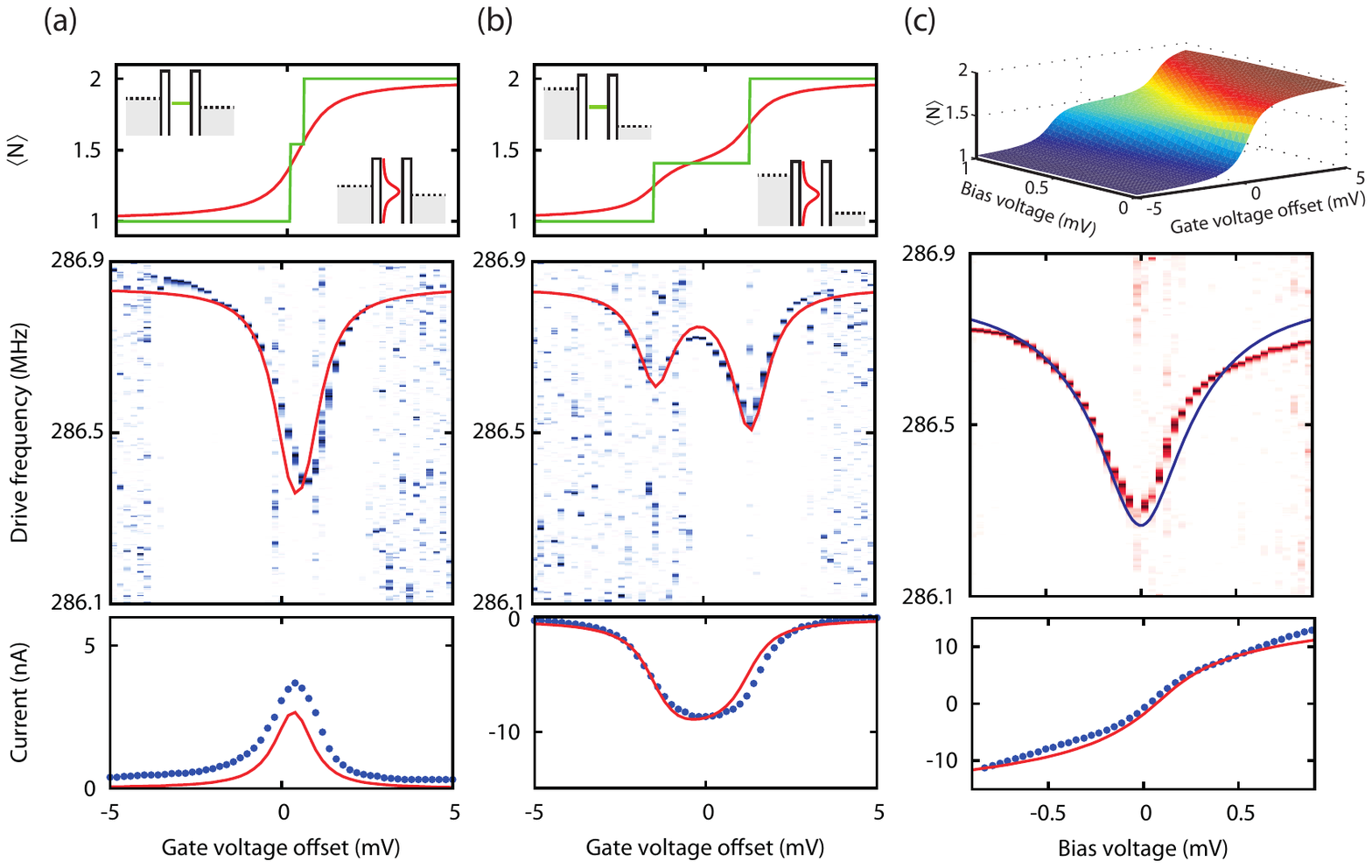} 
	\caption{
(a, top) Calculated average occupation, $\langle N\rangle$, as a function of gate voltage at $V_b = 0.17$ mV, showing two discrete steps for $\Gamma_{broad} =0$ (green line), and a smeared-out monotonic increase for $\Gamma_{broad} = 450$ GHz (red line). Insets: energy diagrams at low bias for zero broadening (left) and finite broadening (right). (a, b, c, middle) Measured normalized current, $|(I-I_{0})|/|I-I_0|_{max}$, as a function of drive frequency and (a, blue) gate voltage offset at $V_b = 0.17$ mV (b, blue) gate voltage offset at $V_b=-0.55$ mV, and (c, red) bias voltage at a gate voltage offset $V_{g,offset} = 0.04$ mV and the corresponding calculated resonance frequency (red line, red line, blue line), showing (a) a decrease in resonance frequency around $V_{g,offset}=0$, (b) a double dip in resonance frequency, (c) a decrease in resonance frequency for low bias voltages. (a, b, c, bottom) Measured (blue dots) and calculated (red line) current as a function of gate voltage at (a) $V_b = 0.17$ mV, (b) $V_b = -0.55$ mV, and (c) as a function of bias voltage at $V_{g,offset} = 0.04$ mV. (b, top) Calculated average occupation, $\langle N\rangle$, as a function of gate voltage at $V_b = -0.55$ mV, showing two discrete steps for both $\Gamma_{broad} =0$ (green line) and for $\Gamma_{broad} = 450$ GHz (red line). Insets: energy diagrams at high bias for zero broadening (left) and finite broadening (right). (c, top) Calculated average
occupation, $\langle N\rangle$, as a function of gate voltage and bias voltage at $\Gamma_{broad} = 450$ GHz, showing the average occupation, $\langle N\rangle$, from increasing in a single step at zero bias to increasing in two steps at $V_b = 1$ mV. 
}
\label{fig2}
\end{figure*}

The calculated average occupation, at a bias voltage of $V_b = 0.17$ mV, corresponding to the white dashed line labeled (b) in Fig. \ref{fig2new}a, is shown in the top panel of Fig. \ref{fig2}a as a function of gate voltage at the charge transition from 1 to 2 electrons. At zero broadening, $\Gamma_{broad}=0$, the green line shows the average occupation increase in two discrete steps. The left inset of Fig. \ref{fig2}a shows an energy diagram illustrating zero broadening. The three discrete plateaus in average charge correspond to Coulomb blockade, single-electron tunneling, and Coulomb blockade again. At a broadening of $\Gamma_{broad}=450$ GHz, for which $\hbar\Gamma_{broad}>e|V_b|$, the red line shows the average occupation increases monotonically in a single step. The right inset of Fig. \ref{fig2}a illustrates how the double step in average occupation, seen at zero broadening, is completely smeared out. Here, $e$ is the elementary charge, $k_B$ is Boltzmann's constant, and $\hbar$ is the reduced Planck's constant.

At a bias voltage of $V_b=-0.55$ mV, corresponding to the white dashed line labeled (c) in Fig. \ref{fig2new}a, the calculated average occupation is shown in the top panel of Fig. \ref{fig2}b as a function of gate voltage. Both at a broadening of $\Gamma_{broad}=0$ (green) and at a broadening of $\Gamma_{broad}=450$ GHz (red), two steps and three plateaus are visible. The insets of the top panel of Fig. \ref{fig2}b show how all allowed states on the CNT are inside the bias window, both for zero broadening (left) and finite broadening (right). Contrary to low bias, the two steps are not smeared into a single step at finite broadening. 

For different bias voltages, the calculated average occupation is shown in the top panel of Fig. \ref{fig2}c as a function of gate voltage for different bias voltages. At low bias, the charge transition takes place in a small range in gate voltage. This gives rise to a large slope for the average occupation with respect to gate voltage. As the bias voltage is increased, the range in gate voltage increases. When the bias voltage is larger than the broadening, the average occupation increases in two steps.

\subsection{Model for the single-electron spring}
\label{subs:modelspring}
Using the average charge as determined in the previous subsection, the shift in resonance frequency is modeled as follows.
The electrostatic force acting on the CNT depends on the voltage difference between the CNT and the gate electrode\cite{Brink2007}:

\begin{equation}
F_{CNT}=\frac{1}{2}\frac{dC_g}{dx}(V_g-V_{CNT})^2,
\label{fig:FCNT}
\end{equation}
where $C_g$ is the capacitance between the CNT and the gate, $x$ is the displacement of the fundamental mode of the CNT, and $V_g$ and $V_{CNT}$ are the voltages on the gate and the CNT, respectively. The voltage on the CNT is determined by the control charge, $q_c=C_gV_g+C_SV_S+C_DV_D$, which is the charge that would be on the CNT in the absence of Coulomb blockade, and the average occupation, $\langle N\rangle$, of charges residing on the CNT:
\begin{equation}
V_{CNT}= \frac{q_c-e\langle N\rangle}{C_{tot}}.
\label{eq:VCNT}
\end{equation}
Here, $C_{S,D}$ and $V_{S,D}$ are the capacitances to and the voltages on the source and drain respectively, and $C_{tot}=C_g+C_S+C_D$. In our case, the bias voltage is applied to the source, $V_S=V_b$, and the drain electrode is grounded, $V_D=0$. Because the charge $N$ on the dot fluctuates stochastically between $N_0$ and $N_0+1$ at a rate $\Gamma_{tot} \gg f_0$, the mechanical motion experiences a voltage on the CNT due to an average occupation, $\langle N\rangle$.

The spring constant and, consequently, the resonance frequency of the CNT are determined by the change in force acting on the CNT per unit displacement:
\begin{equation}
2m\omega_0\Delta\omega_0=\Delta k=-\frac{dF}{dx},
\label{eq:springconst}
\end{equation}
where $m$ is the mass of the CNT, and $\Delta \omega_0\ll \omega_0$. The full derivative of the force with respect to the displacement of the CNT is expanded into partial derivatives with respect to displacement and gate voltage:
\begin{align}
-\frac{dF(x,V_g)}{dx}=&-\frac{\partial F(x,V_g)}{\partial x}-\frac{dC_g}{dx} \frac{dq_c}{dC_g} \frac{dV_g}{dq_c} \frac{\partial F(x,V_g)}{\partial V_g}\nonumber\\
=&-\frac{1}{2}\frac{d^2C_g}{dx^2}(V_g-V_{CNT})^2\nonumber\\
&-\frac{V_g(V_g-V_{CNT})}{C_g}\left(\frac{dC_g}{dx}\right)^2\frac{\partial(V_g-V_{CNT})}{\partial V_g}.
\label{eq:dFdx}
\end{align}
The first term gives rise to a softening spring effect due to the capacitive force between the CNT and the gate. The second term takes into account the influence of the displacement on the control charge through the gate capacitance. In turn, the influence of the control charge on the force is incorporated through the gate voltage. In this article, we focus on changes in the spring constant that occur rapidly with gate voltage and we show that the experimental features can be captured using only the second term.

Combining Eq. (\ref{eq:VCNT}) and the second term of (\ref{eq:dFdx}) leads to the following expression for the change in resonance frequency due to a changing average charge\cite{Steele2009}:
\begin{equation}
	\Delta \omega_0=\frac{V_g(V_g - V_{CNT})}{2m\omega_0 C_{tot}}\left(\frac{dC_g}{dx}\right)^2\left(1-\frac{C_{tot}}{C_{g}}-\frac{e}{C_{g}}\frac{\partial\left\langle N \right\rangle}{\partial V_g}\right).
\label{eq:deltaomega}
\end{equation}
Because of the rightmost minus sign and the fact that $\partial \left\langle N \right\rangle/\partial V_g>0$, the changing average charge on the CNT leads to a softening spring effect. The resonance frequency of the CNT decreases more when the mechanical oscillation causes a larger change in average charge, expressed by $\partial \left\langle N \right\rangle/\partial V_g$.

\subsection{Fitting of the experimental resonance frequency shift and discussion}
\label{subs:fitspring}
To verify the model established in the previous two paragraphs, we perform a quantitative fit on the experimental data shown earlier in Fig. \ref{fig2new}. The fits are accomplished by using Eq. \ref{eq:deltaomega} and a single set of parameters for all figures. The values chosen for the parameters can be found in appendix \ref{app:param}. A value for $\partial\left\langle N \right\rangle/\partial V_g$ is obtained by numerically differentiating the calculated average occupation, as is displayed in the top panel of Fig. \ref{fig2}a, with respect to gate voltage. 

The red graphs in the middle panel of Fig. \ref{fig2}a and b show the calculated change in resonance frequency as a function of gate voltage. At $V_b=0.17$ mV, Fig. \ref{fig2}a shows quantitative agreement between the measurement and the model describing the frequency dips. The frequency dip corresponds to the largest slope of the average charge with gate voltage. At $V_b=-0.55$ mV, Fig. \ref{fig2}b demonstrates how the model reproduces the experimentally observed double dip structure. At the plateau in average charge, the mechanical oscillation only brings about a small change in average charge, and the resonance frequency returns towards its original value.

To investigate the reduction in the resonance frequency as a function of bias voltage, a vertical linecut is taken, at a constant gate voltage through the charge degeneracy point in the stability diagram. The middle panel of Fig. \ref{fig2}c shows the measured current, but now as a function of bias voltage, not gate voltage, and drive frequency. The calculated resonance frequency in blue shows excellent quantitative agreement with the measurement. At low bias voltages, the narrow charge transition leads to a large change in average charge due to the mechanical oscillation and a large change in the displacement-dependent force, resulting in a large decrease in resonance frequency. At higher, both positive and negative, bias voltage, the slope in average occupation, $\partial\langle N\rangle/\partial V_g$, becomes less, as the plateau arises. This leads to a smaller change in the displacement-dependent force and a smaller decrease in the resonance frequency.

The experimentally obtained current flowing through the CNT is shown with blue dots in the bottom panels of Fig. \ref{fig2} at $V_b=0.17$ mV (a), $V_b=-0.55$ mV (b), and as a function of bias voltage (c). For $V_b=0.17$ mV, a Coulomb peak is visible. For $V_b=-0.55$ mV, the plateau in current shows that the bias voltage, $V_b$, is larger than the broadening, $\Gamma_{broad}$. As a function of bias, the bottom panel of Fig. \ref{fig2}c shows no Coulomb blockade, since the vertical linecut in bias voltage exactly passes through the charge degeneracy point. The absence of a saturation of the current at high bias voltage indicates low tunnel barrier heights and supports the choice of energy-dependent tunnel rates. As an independent examination of the chosen tunnel rates, the measured current is fitted using the same parameters as for the frequency dips (red lines). The current is calculated using:
\begin{equation}
I = e\frac{\Gamma^+_L\Gamma^-_R-\Gamma^+_R\Gamma^-_L}{\Gamma_{tot}}.
\label{eq:current}
\end{equation}

\begin{figure*}[ht]
\centering
	\includegraphics{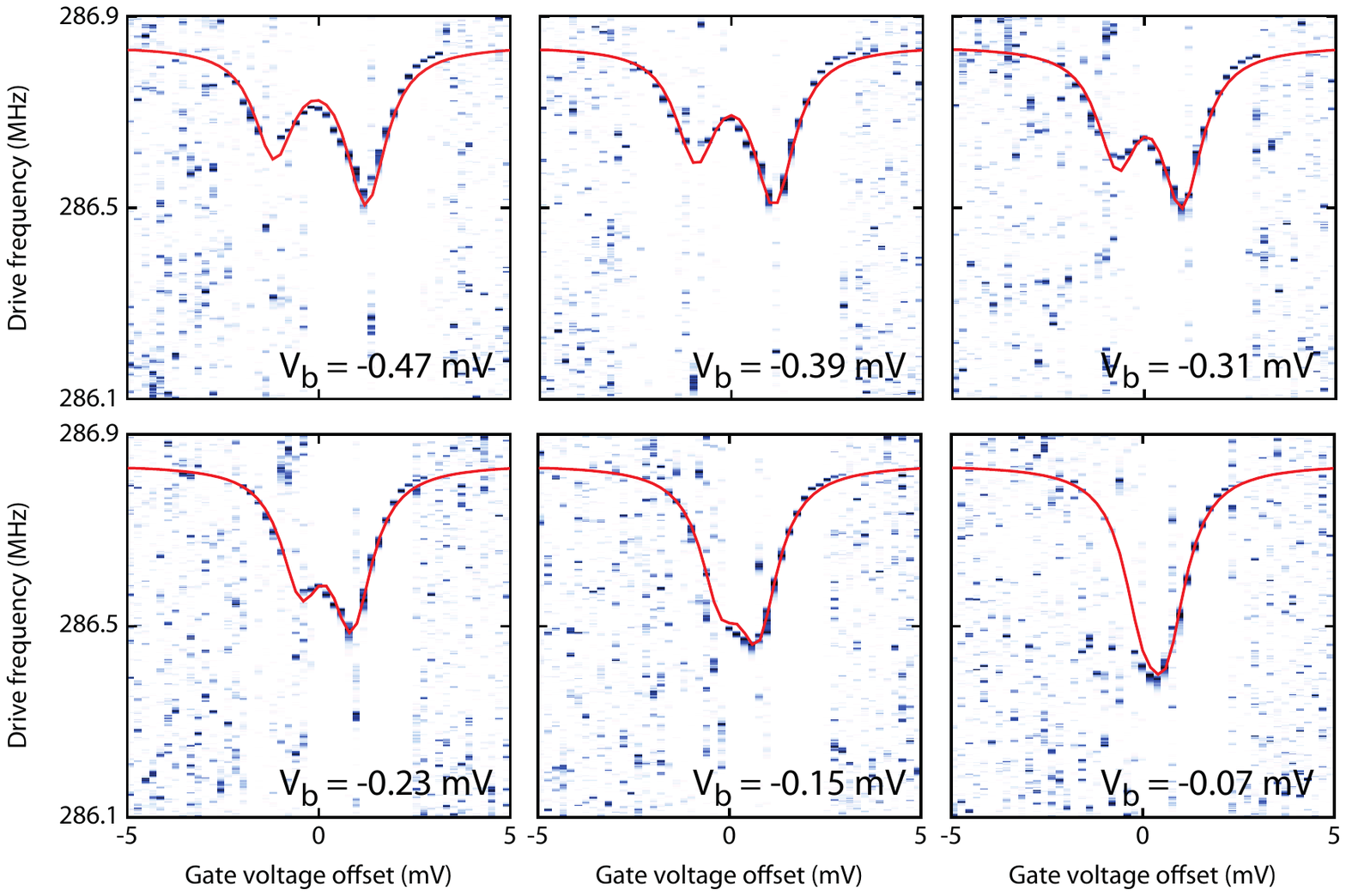} 
	\caption{Measured normalized current (blue) and calculated resonance frequency (red line) as a function of gate voltage offset for different bias voltages, showing the transition of the double dip structure into a single dip as the bias voltage becomes less negative.}
\label{fig:negbiasfreqfits}
\end{figure*}

For intermediate bias voltages, Fig. \ref{fig:negbiasfreqfits} illustrates the transition from the double dip structure to a single dip. Using the same set of parameters as for Fig. \ref{fig2}, the calculated resonance frequency follows the measurement well. The quantitative agreement between the experiment and the calculation of the resonance frequency and the current for different bias voltages demonstrates the consistency of the single set of parameters used for the calculation.


\section{Single-electron damping}
\label{sec:SEdamping}
In this section, the influence of the changing average charge on the mechanical quality factor of the CNT is investigated. As with the resonance frequency in the previous section, the quality factor is determined experimentally and by using a model, for low bias voltage, high bias voltage, and as a function of bias voltage. To show the consistency of the used parameters, the corresponding resonance frequency is also displayed.

Figure \ref{fig3}a shows the stability diagram of the charge transition from 7 to 8 electrons at a parallel magnetic field of 8 T. The current flowing through the CNT is measured as a function of bias and gate voltage, and the derivative $dI/dV_b$ is determined numerically. The letters b and d denote, respectively, the low and high bias voltage, at which Figs. \ref{fig3}b and \ref{fig3}d were taken. The letter f denotes the gate voltage at which Fig. \ref{fig3}f was taken.

\subsection{Experimental observation of single-electron damping}

The experimentally determined mechanical quality factor at a low bias voltage of $V_b=0.3$ mV is displayed in blue dots in the top panel of Fig. \ref{fig3}b. The quality factor is obtained by measuring the current versus drive frequency 10 times and averaging. The drive power is adjusted for each quality factor measurement such that the frequency response shows a lineshape which is as close as possible to Lorentzian, but which is not obscured by noise. By working at low powers in the linear response regime, we minimize the likelihood that non-linear damping terms\cite{Eichler2011} play a significant role. The change in current, $\Delta I$, due to the mechanical motion in our detection scheme is proportional to the amplitude of the displacement squared. At the sides of the Coulomb peak, the lack of nonlinearity of the current with gate voltage prevents the measurement of the mechanical motion through a change in current, and consequently a determination of the quality factor. A fit with the following Lorentzian function is performed to extract the quality factor.
\begin{equation}
\Delta I(\omega) = {\cal A}\frac{\omega_0^4/Q^2}{((\omega_0^2-\omega^2)^2+\frac{\omega_0^2\omega^2}{Q^2})}.
\label{eq:fitresonance}
\end{equation}
Here, $Q$ is the mechanical quality factor, $\omega$ is the drive frequency, and ${\cal A}=\frac{1}{4}\left(\frac{V_g}{C_g}\frac{dC_g}{dx}\right)^2 \frac{d^2I}{dV_g^2} x^2$ is a fit parameter, which incorporates the amplitude of oscillation $x$ and the electromechanical coupling. At a low bias of $V_b=0.3$ mV, a large decrease in the quality factor of two orders of magnitude is visible, spanning several millivolts of gate voltage corresponding to single-electron tunneling. The top panel of Fig. \ref{fig3}d shows the experimentally obtained quality factor in blue dots at a high bias voltage of $V_b=1.1$ mV. Here, similarly to the double frequency dip at high bias in the middle panel of Fig. \ref{fig2}b, two dips in quality factor are visible.

To investigate the relation between current and single-electron damping of the mechanical motion, Fig. \ref{fig3}e shows the current as a function of bias voltage. The gate voltage is chosen such that the graph is a vertical linecut through the charge degeneracy point. The top panel of Fig. \ref{fig3}f shows the corresponding quality factor as a function of bias voltage, again at a gate voltage of $V_g=0.9002$ V. The quality factor exhibits a dip at low bias voltages and returns to its original value at higher, both negative and positive, bias voltage. We emphasize that the highest current flowing through the CNT does not correspond to the largest reduction in quality factor. Instead, in fact, when the gate voltage is positioned at a Coulomb peak, the quality factor actually goes up when increasing the current through the device.

Figure \ref{fig3}c shows the basic concept of the damping mechanism, illustrated with zero broadening and one electron tunneling event during a mechanical oscillation. Let us first consider the zero bias voltage case. The mechanical motion brings the chemical potential of the CNT below and above the Fermi level of the leads. The asymmetry of the Fermi sea in the leads implies that electrons can only tunnel onto the CNT when the level of the CNT is below the Fermi energy of the leads, and can only tunnel off when it is above. As the mechanical oscillation pushes the level downwards through the Fermi energy, there is a small retardation in the time when the electron tunnels on, given by $1/\Gamma$. As the level is pushed back upwards in electrostatic energy by the motion, the electron tunnels off, again with a small time delay. The net result is that electrons are pumped from below the Fermi energy to above it, extracting energy from the mechanical motion, and thus resulting in mechanical damping. Although the retardation time is small compared to the mechanical frequency ($\Gamma \gg \omega_0$), the damping is still large due to the large electro-mechanical coupling. At finite bias voltage, the same picture applies, but now the damping occurs when the level of the CNT passes by the Fermi energy of each of the left and right lead separately. 

At zero temperature and with no broadening of the transition from the tunnel coupling to the leads, the dip in quality factor would be infinitely narrow. Including finite temperature and tunnel coupling of the level to the leads, the dip in quality factor acquires a finite width, and is proportional to the change in average occupation with gate voltage, $\partial\langle N\rangle/\partial V_g$, as shown in Eq. \ref{eq:damping} in the next subsection. The fact that the quality factor is determined in this way also explains the observed single and double dip structures in the quality factor. At low bias voltage, the charge transition takes place in a single step and consequently there is a single dip in the quality factor. As the bias voltage is increased, the single step changes into two steps with a plateau between them. On the plateau, the change in average occupation with gate voltage, $\partial\langle N\rangle/\partial V_g$, is smaller, leading to a smaller retarded single-electron force acting on the CNT and a smaller reduction of the quality factor. Although a larger current is due to more tunnel events, the CNT does not perform extra work at high bias. As more electrons pass the CNT and reside on it, the retardation time is reduced. What causes the reduction in quality factor is the pumping of the electrons from below the Fermi energy to above it, and the retardation between the electrostatic force and the mechanical motion.

\subsection{Model for single-electron damping}

\begin{figure*}[ht]
\centering
	\includegraphics{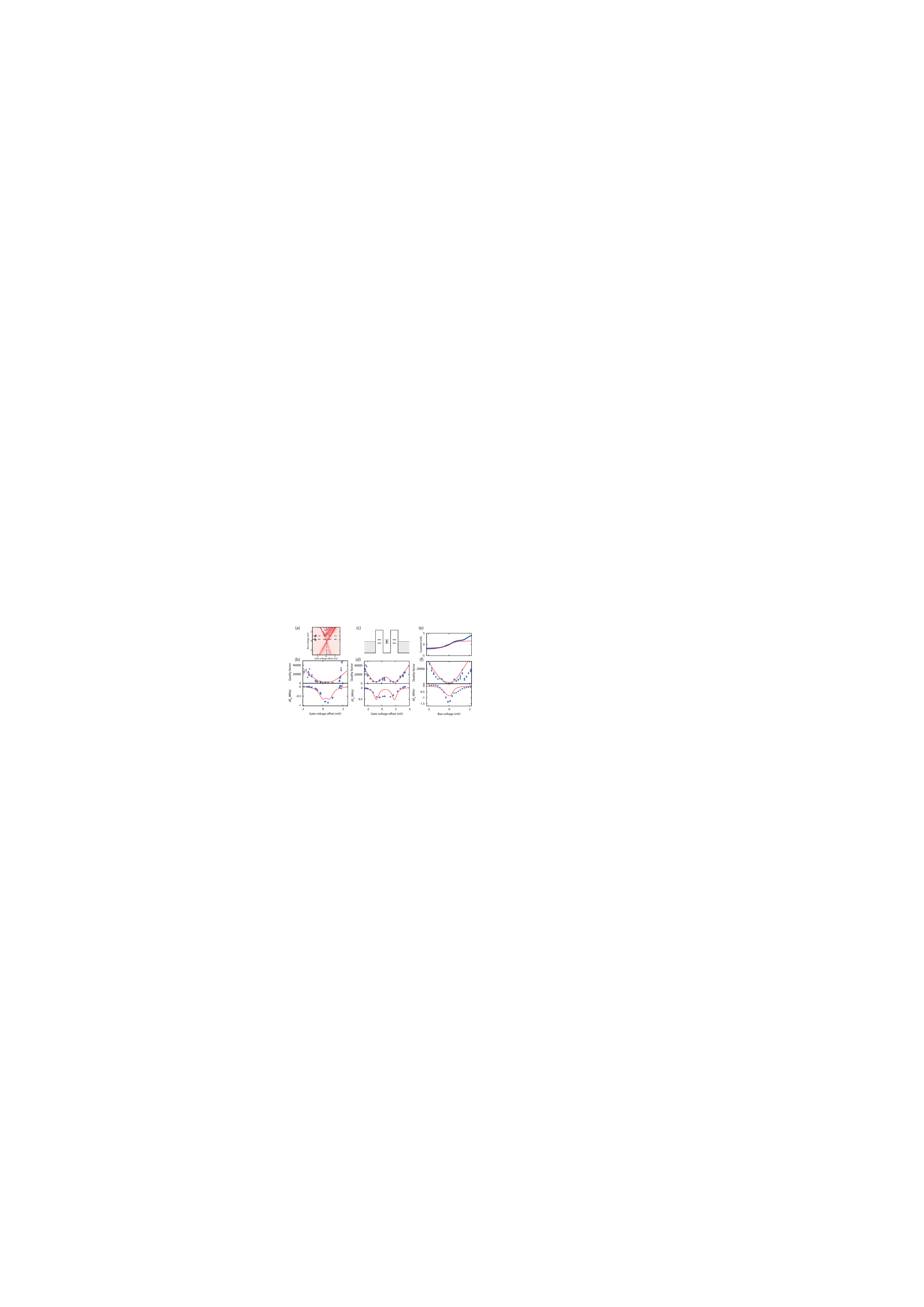}
	\caption{(a) Stability diagram at a parallel magnetic field of 8 T and $V_g = 0.9$ V, showing the charge transition from 7 to 8 electrons, with 
the gate and bias voltages at which Figs. \ref{fig3}b, \ref{fig3}d, and \ref{fig3}f were taken, indicated by black and blue dashed lines. (b, d, f) Comparison between measurement (blue dots) and model (red lines) of the quality factor (top) and the
shift in resonance frequency (bottom) as a function of gate voltage (b, d) and bias voltage (f), showing, (b) at $V_b = 0.3$ a single dip, (d) at $V_b = 1.1$ mV, a double dip, and (f) at $V_{g,offset} = 0$ mV, a large decrease in quality factor and resonance frequency at low bias voltages. (c) Energy diagram (exaggerated) illustrating the damping mechanism, showing the asymmetry of electrons tunneling onto the CNT at low chemical potential and off the CNT at high chemical potential. (e) Measured (blue dots) and calculated (red line) current as a function of bias voltage at $V_{g,offset} = 0$ V.}

\label{fig3}

\end{figure*}

Using the following model, the quality factor is calculated, as shown in the red lines in the top panels of Figs. \ref{fig3}b, \ref{fig3}d, and \ref{fig3}f. We look at the limit where there are many tunneling events per mechanical oscillation, $\Gamma_{tot} \gg f_0$. As derived in Refs.~\onlinecite{Usmani2007,Labadze2011}, the total damping has an intrinsic contribution and a contribution due to the displacement-dependent force associated with tunneling electrons:
\begin{equation}
\frac{\omega_0}{Q_{tot}}=\frac{\omega_0}{Q_{int}}+\frac{F_{stoch}V_g}{mC_g}\frac{1}{\Gamma_{tot}}\frac{dC_g}{dx}\frac{\partial\langle N\rangle}{\partial V_g},
\label{eq:damping}
\end{equation}
where $Q_{tot}$ is the total quality factor and $Q_{int}$ is the intrinsic quality factor in the absence of tunneling electrons. The stochastic force experienced by the CNT, $F_{stoch}=F(N_0+1)-F(N_0)$, is the difference between the force experienced at $N_0$ and $N_0+1$ electrons. Assuming that both source and drain voltage are much smaller than the d.c. gate voltage, $V_s, V_d\ll V_g^{dc}$, and taking only into account the d.c. force acting on the electrons as the CNT oscillates (which is valid as long as $V_g^{ac}\ll V_g^{dc}$), Ref.~\onlinecite{Labadze2011} gives the following expression for the stochastic force:
\begin{equation}
	F_{stoch}=\frac{1}{C_{tot}^2}\frac{d C_g}{dx}\left(2e(C_S+C_D) V_g+e^2(2N+1)\right).
	\label{eq:Fstoch}
\end{equation}
Equation (\ref{eq:damping}) illustrates how single-electron damping is increased by a large change in average occupation because of mechanical motion, $\partial\langle N\rangle/\partial V_g$, but is reduced by a large total tunnel rate, $\Gamma_{tot}$, as the retardation time decreases. In the case that exponential tunnel rates cause $\partial\langle N\rangle/\partial V_g$ to become negative, single-electron tunneling pumps the mechanical motion, leading to self-sustained oscillation\cite{Usmani2007}.

\subsection{Fitting of the experimental quality factor and discussion}

The calculated and the measured quality factors, at low bias, are compared in the top panel of Fig. \ref{fig3}b, showing quantitative agreement below $V_{g,offset}=1$ mV, where $V_{g,offset}=V_g-0.9$ V. At high bias, the top panel of Fig. \ref{fig3}d shows an excellent quantitative agreement between the model and the measurement. As a function of bias in the top panel of Fig. \ref{fig3}f, the calculated quality factors match the measured quality factors at bias voltages below $1$ mV. Above $1$ mV, an excited state, as evident in the current in Fig. \ref{fig3}e, may be the cause of extra damping.

The bottom panels of Figs. \ref{fig3}b, \ref{fig3}d, and \ref{fig3}f show, in blue dots, the resonance frequency extracted from the frequency responses alongside the quality factors. The decrease in quality factor is accompanied by a decrease in the resonance frequency. The red lines show the resonance frequency calculated with the model as described in the previous section, using the same tunnel rates as for the quality factors. Without changing any fit parameters, we have qualitative agreement as the quality factor dips and the resonance frequency dips occur at corresponding gate voltages. It is not fully understood why agreement between the calculated and measured resonance frequency is not quantitative.

\section{Nonlinear restoring forces due to single-electron tunneling}
\label{sec:SEnonlinearity}
In the previous sections, we described linear corrections to the restoring force due to single-electron tunneling, which change the spring constant and cause damping. In this section, we cover nonlinear corrections due to single-electron charge effects\cite{Meerwaldt2012}, resulting in expressions for the Duffing parameter and the mode coupling parameter.

\subsection{Duffing nonlinearity due to single-electron tunneling}
\label{subs:SEDuffing}
Adding nonlinear corrections, the single-electron force is expanded with respect to the displacement, $x$, as follows:
\begin{equation}
F_{CNT} = -\Delta k_{SET}x-\beta_{SET}x^2-\alpha_{SET}x^3.
\label{eq:FSET}
\end{equation}
The $\beta_{SET}$ parameter and the Duffing parameter, $\alpha_{SET}$, are calculated as derivatives of the single-electron force:
\begin{align}
\beta_{SET}=-\frac{1}{2}\frac{d^2F_{CNT}}{dx^2}\\
\alpha_{SET}=-\frac{1}{6}\frac{d^3F_{CNT}}{dx^3}.
\label{eq:alphaandbeta2}
\end{align}
Expanding the full derivatives into partial partial derivatives as $d/dx=\partial/\partial x+(V_g/C_g)(dC_g/dx)\partial/\partial V_g$, we arrive at:
\begin{align}
\beta_{SET}=&-\frac{1}{2}F_{xx}-\frac{V_g}{C_g}\frac{d C_g}{dx}F_{xg}-\frac{1}{2}\left(\frac{V_g}{C_g}\frac{d C_g}{dx}\right)^{2}F_{gg}\\
\alpha_{SET}=&-\frac{1}{6}F_{xxx}-\frac{1}{2}\frac{V_g}{C_g}\frac{d C_g}{dx}F_{xxg}\nonumber\\
&-\frac{1}{2}\left(\frac{V_g}{C_g}\frac{d C_g}{dx}\right)^{2}F_{xgg}
-\frac{1}{6}\left(\frac{V_g}{C_g}\frac{d C_g}{dx}\right)^{3}F_{ggg}.
\label{eq:alphaandbeta}
\end{align}
Here, the subscripts of $F$ denote differentiation with respect to displacement, $x$, or gate voltage, $g$. 

\begin{figure}[ht]
\centering
	\includegraphics{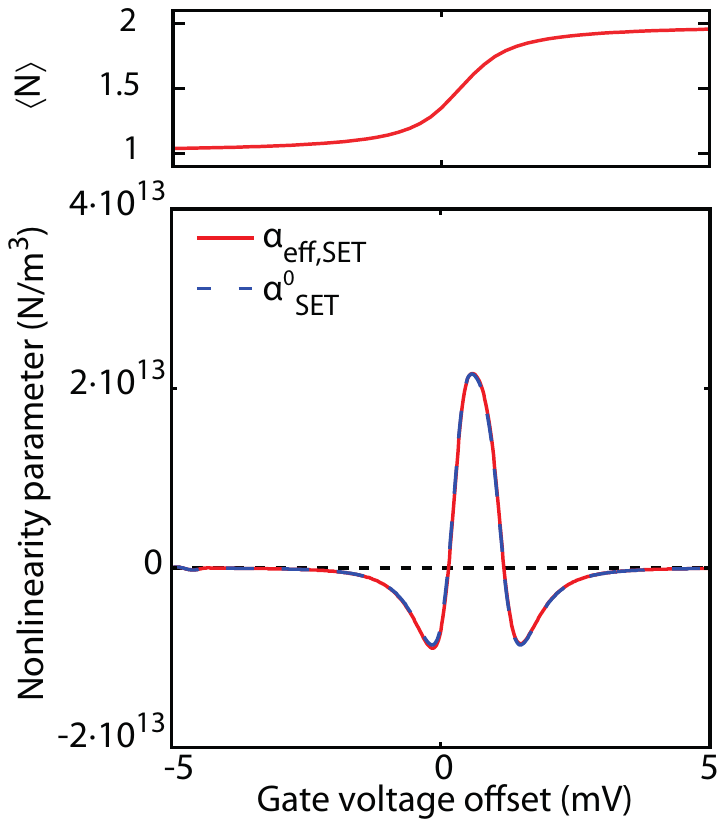}
	\caption{(top) Average occupation, $\langle N\rangle$, as a function of gate voltage, identical to the top panel of Fig. \ref{fig2}a. (bottom) Calculated nonlinearity parameter as a function of gate voltage at $V_b= 0.17$ mV across the Coulomb peak depicted in Fig. \ref{fig2new}, with its sign flipping from negative to positive to negative again, showing the significance of the $\alpha^0_{SET}$ contribution in $\alpha_{eff,SET}$. Other contributions, including the renormalization, $\Delta\alpha_{\beta,SET}$, due to $\beta_{SET}$, are shown to be negligible. 
}
\label{fig:comparealphas}
\end{figure}

The quadratic nonlinearity, $\beta_{SET}$, renormalizes\cite{Lifshitz2008} $\alpha_{SET}$, leading to an effective Duffing parameter, $\alpha_{eff, SET}$:
\begin{equation}
\alpha_{eff, SET}=\alpha_{SET}+\Delta\alpha_{\beta,SET}=\alpha_{SET}-\frac{10}{9}\frac{\beta^2_{SET}}{m \omega_0^2}.
\end{equation}
For the parameters of the device we study here, there is a leading order dominant contribution, $\alpha^0_{SET}$, given by:
\begin{equation}
\alpha^0_{SET}=-\frac{1}{6}\left(\frac{dC_g}{dx}\right)^{4}\left(\frac{V_g}{C_g}\right)^{3}\frac{e(V_g-V_{CNT})}{C_{tot}}\frac{\partial^3\langle N\rangle}{\partial V_g^3},
\label{eq:alpha0}
\end{equation}
which is one of the terms arising from $-1/6((V_g/C_g)(dC_g/dx))^3F_{ggg}$ in Eq. \ref{eq:alphaandbeta}. Using the same set of parameters as for Figs. \ref{fig2} and \ref{fig:negbiasfreqfits}, we repeat the calculated average occupation as displayed in Fig. \ref{fig2}a, and we plot $\alpha_{eff,SET}$ and $\alpha^0_{SET}$ together in Fig. \ref{fig:comparealphas}. The figure shows that the other contributions in Eq. \ref{eq:alphaandbeta} and the renormalization, $\Delta\alpha_{\beta,SET}$, due to $\beta_{SET}$, play no significant role in our device. For completeness, the other contributions, which are more than one order of magnitude smaller than $\alpha^0_{SET}$, are plotted in Fig. \ref{fig:appothercontributions} in appendix \ref{app:othercontributions}. The switching of the sign of the Duffing parameter $\alpha_{eff,SET}$, due to the third derivative of the average occupation with respect to gate voltage, is visible in Fig. \ref{fig:comparealphas} and has been observed experimentally previously\cite{Steele2009}.

\subsection{Mode coupling due to single-electron tunneling}

Recently\cite{Castellanos-Gomez2012}, different bending modes in CNT quantum dot resonators have experimentally been shown to have strong coupling, resulting in a shift in the resonance frequency of one mode due to the resonance of another mode. In contrast to top-down micromechanical beams\cite{Westra2010} and CNTs outside the Coulomb blockade regime\cite{Eichler2012}, mode coupling in CNT quantum dot resonators is not dominated by tension, but, instead, by single-electron charge effects. In this section, we establish a theoretical framework for single-electron mode coupling. We focus on the leading order nonlinear contribution to the restoring force, arising from $F^0_{SET}=-\alpha^0_{SET}x^3$, and write it in terms of the change in capacitance, $\delta C_g$, due to mechanical motion:
\begin{equation}
F^0_{SET}=-\mathcal{B}(V_g)\frac{\partial^3\langle N\rangle}{\partial V_g^3}\frac{dC_g}{dx}(\delta C_g)^3,
\end{equation}
where, for brevity, we capture the slowly varying dependence on gate voltage in $\mathcal{B}(V_g)=\frac{1}{6}\left(\frac{V_g}{C_g}\right)^{3}\frac{e(V_g-V_{CNT})}{C_{tot}}$. The change in capacitance due to mechanical motion depends on the mode shape:
\begin{equation}
\delta C_g=\int_0^L{\frac{dc_g}{du}\left(\sum_n{u_n(z)\cos(\omega_nt)}\right)}dz,
\end{equation}
where $dc_g/du$ is the change in capacitance per unit length with respect to the amplitude $u(z)$ at position $z$ along the CNT. We use the infinite cylinder parallel to a plate model\cite{Sapmaz2003}, and assume that $dc_g/du$ is independent of the position, $z$, along the CNT. The position-dependent amplitude, $u_n(z)$ is given by:
\begin{align}
u_n(z) = x_n\xi_n(z),
\end{align}
where the mode shapes, $\xi_n(z)$, are orthonormalized as $\int_0^L{\xi_i(z)\xi_j(z)}dz=L\delta_{ij}$ with $\delta_{ij}$ the Kronecker delta, such that the amplitude $x$ represents the root-mean-squared amplitude over the length of the CNT (not over time). With this definition of the amplitude\cite{Poot2012}, the mass of each mode is given by the total mass and the spring constant is given by $k_n=m \omega_n^2$. The change in gate capacitance due to the mechanical motion is simplified to:
\begin{equation}
\delta C_g=L\frac{dc_g}{du}\sum_n{x_na_n\cos(\omega_nt)},
\end{equation}
with the first four elements of $a_n =(1/L) \int_0^L{\xi_n(z)}dz$ calculated\cite{Poot2012} for a bending-rigidity dominated beam to be 0.83, 0, 0.36, and 0. The leading order nonlinear term, $F^0_{SET}$ is expanded as:
\begin{align}
F^0_{SET}=&-\mathcal{B}(V_g)\frac{\partial^3\langle N\rangle}{\partial V_g^3}\mathcal{C}(\{x_i\})\nonumber\\
&\times\sum_{m,n}{a_m^2x_m^2\left(\frac{1}{2}+\frac{1}{2}\cos(2\omega_mt)\right)a_nx_n\cos{\omega_nt}}
\end{align}
where we have used $\mathcal{C}(\{x_i\})=(Ldc_g/du)^4(\sum_i{a_ix_i})/\sum{x_i}$. 

We address several terms in this expression. The terms containing $(1/2)\cos(2\omega_mt)$ give rise to parametric excitation due to single-electron effects. The terms not containing $(1/2)\cos(2\omega_mt)$, while having $m=n$, lead to the Duffing nonlinearity of mode 1 and 3, as described in the previous subsection. Single-electron mode coupling is brought about through the terms not containing $(1/2)\cos(2\omega_mt)$, while having $m\neq n$, as the amplitude of one mode changes the spring constant of another mode. The change in resonance frequency of mode $m$ due to the resonance of mode $n$ is then given by:
\begin{align}
\frac{\Delta\omega_m}{x_n^2}=-\frac{1}{4m\omega_m}\mathcal{B}(V_g)\frac{\partial^3\langle N\rangle}{\partial V_g^3}\mathcal{C}(x_m,x_n)a_n^2a_m.
\end{align}
As the sign of $\partial^3\langle N\rangle/\partial V_g^3$ goes from positive, to negative, to positive again, across a Coulomb peak, mode coupling causes softening, then stiffening, and then softening, in the CNT spring. The change in sign from mode coupling across a Coulomb peak arises from the mechanically modulated average charge in the same way as the sign of the Duffing nonlinearity does.

\section{Additional experimental observations of single-electron spring and damping effects}
\label{sec:additional}
In sections \ref{sec:SEspring} and \ref{sec:SEdamping}, we showed how the resonance frequency and the quality factor decrease as the average charge on the CNT is modulated through mechanical oscillation. How much the average charge changes is determined by $\partial\langle N\rangle/\partial V_g$, which in turn depends on the broadening $\Gamma_{broad}$ due to tunnel coupling to the leads. First, in subsection \ref{subs:magneticfield}, we show that, by tuning the tunnel rates through magnetic field, the dips in resonance frequency and quality factor are influenced significantly, and in a way that is in qualitative agreement with the physical picture of the damping and frequency shifts from our model. In subsection \ref{subs:excitedstate}, a step in the average charge caused by an excited state of the CNT is presented to lead to a decrease in the resonance frequency, demonstrating detection of the excited state of the quantum dot using the mechanical resonator. Finally, in subsection \ref{subs:FabryPerot}, we demonstrate that the single electron spring and damping effects we observe are indeed originating from Coulomb blockade, by studying the gate dependence of the resonance frequency in the Fabry-P\'{e}rot conductance regime. In this regime, the charge on the suspended CNT segment is no longer quantized, and we no longer observe dips in the mechanical frequency as we sweep the gate.

\subsection{Resonance frequency shifts and damping via magnetic-field dependent tunnel rates}
\label{subs:magneticfield}
In this subsection, we examine the quality factor and mechanical frequency as a function of the tunnel rates of the quantum dot. We tune the tunnel barriers of the quantum dot in a somewhat unconventional way by using a magnetic field parallel to the CNT. 

\begin{figure}[ht]
\centering
	\includegraphics{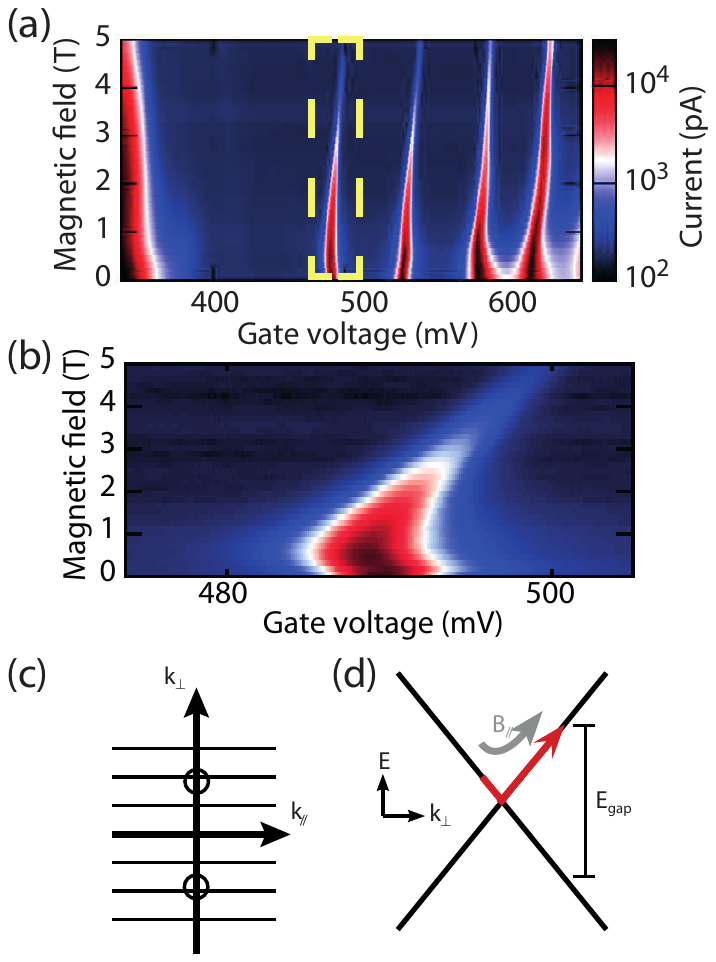}
	\caption{(a) Current as a function of gate voltage and parallel magnetic field, showing a decrease in
the height and width of the Coulomb peaks and a shift of the Coulomb 
peaks with increasing magnetic field for the first four charge states. The yellow dashed box denotes the region shown in \ref{fig4}b. (b) Current as a function of gate voltage and parallel magnetic field, showing the shift of the transition from 0 to 1 electrons with magnetic field (c) Cross-section of the Dirac cones of a CNT at a constant energy as a function of $k_\bot$ and $k_\|$, showing the intersection of the quantized $k_\bot$ with the Dirac cones. (d) Cross-section of a Dirac cone of a CNT, showing the change in the bandgap, $E_{gap}$, as the parallel magnetic field alters $k_\bot$.
}
\label{fig4}
\end{figure}

The tuning of the tunnel barriers by magnetic field occurs through the parallel magnetic field's influence on the bandgap of the CNT. Figure \ref{fig4}a shows the current flowing through the CNT as a function of gate voltage and magnetic field parallel to the CNT. The increase in the bandgap is visible as an increase in the gate voltage range between the hole current and the first Coulomb peak, which can be seen clearly by the trajectory of the first Coulomb peak in Fig. \ref{fig4}b. Also, as the magnetic field is increased, the width and the magnitude of the Coulomb peaks decreases. This decrease in the Coulomb peak width is a result of an increase in the p-n junction tunnel barrier height by the increased bandgap at higher magnetic fields.

The increase of the bandgap with an increasing parallel magnetic field can be explained as follows\cite{Minot2004}. A parallel magnetic field changes the quantized wavevectors $k_\bot$ of the electrons along the circumference of the CNT through an Aharonov-Bohm term. The band structure of the CNT is determined by taking a cross-section of the Dirac cone at a constant $k_\bot$. Figure \ref{fig4}c shows the quantized wavevectors $k_\bot$ intersecting with the Dirac cones. Figure \ref{fig4}d illustrates how the bandgap is changed through a parallel magnetic field. The height of the tunnel barrier is determined by the bandgap of the CNT, through the p-n junction that is formed at the interface of the metal and the CNT. In general, the shift of the quantization lines with magnetic field decreases the bandgap. This occurs only up until a magnetic field $B_{Dirac}$, at which point the quantization line crossing the Dirac point and the bandgap begins to increase again. From the orbital magnetic moment we observe of 1.0 meV/T, together with the bandgap of $E_g = 58$ meV at zero magnetic field, this should occur at a very large magnetic field of 29 T. Similar to previous reports \cite{Deshpande2009}, we observe that the quantization line crosses the Dirac point at a magnetic field much smaller than expected, in this case at 0.6 T. At a magnetic field above $B_{Dirac}$, the p-n junction tunnel barriers to the quantum dot increase in height, and the tunnel rates to the quantum dot are significantly reduced. 

\begin{figure}[ht]
\centering
	\includegraphics{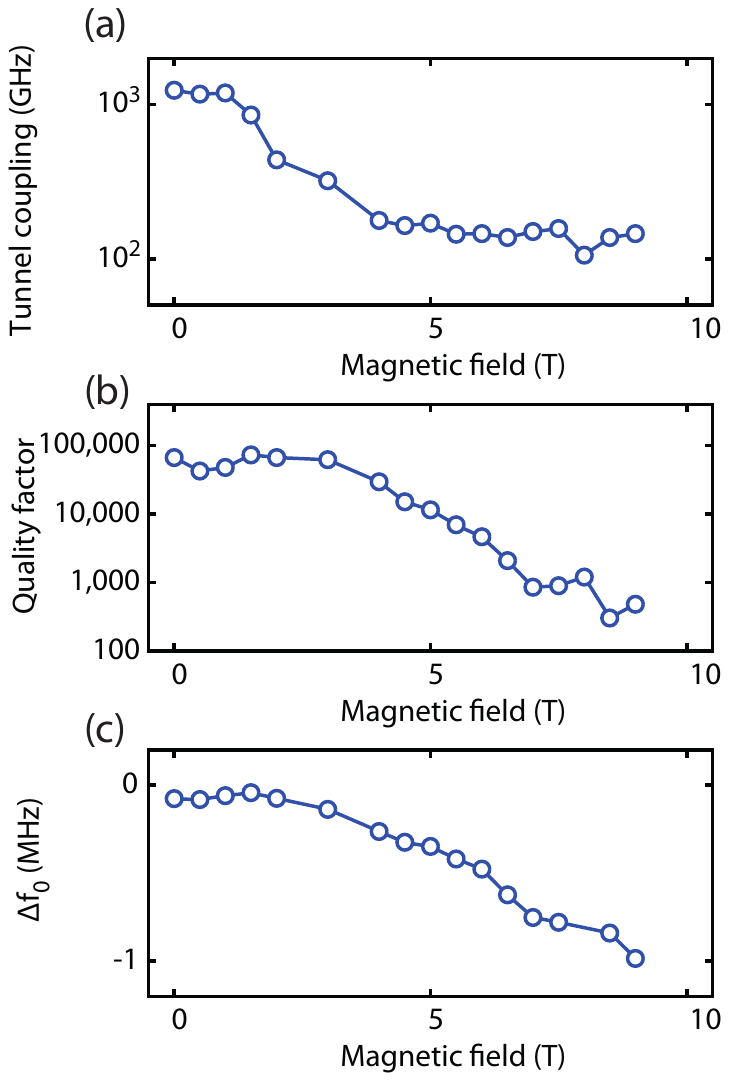}
	\caption{(a) Broadening due to tunnel coupling as a function of magnetic field, showing a decrease in broadening with increasing magnetic field. (b) Quality factor on the Coulomb peak of the 7 to 8 transition at $V_b = 0.3$ mV showing the quality factor decreasing by two orders of magnitude with increasing magnetic 
field. (c) Shift in resonance frequency on the Coulomb peak of the 7 to 8 transition at $V_b = 0.3$ mV, showing a decrease of $\sim$1 MHz.}
\label{fig4new}
\end{figure}

The tunnel rate of the quantum dot to the leads as a function of magnetic field, $\Gamma_{broad}(B)$, is determined from the observed width of the Coulomb peaks. Figure \ref{fig4new}a shows the tunnel coupling as a function of parallel magnetic field, decreasing with an order of magnitude. To determine the tunnel coupling, a Lorentzian fit, similar to Eq. (\ref{eq:broadening}), was performed on the Coulomb peak at $V_b = 0.3$ mV. 

In Figs. \ref{fig4new}b and \ref{fig4new}c, the resonance frequency and quality factor as a function of parallel magnetic field are shown, respectively. The resonance frequency and quality factor are determined by taking a frequency response at the top of the Coulomb peak of the charge transition from 7 to 8 electrons, at a low bias voltage of $V_b=0.3$ mV. Between 0 T and 9 T, the reduction of the tunnel coupling with a parallel magnetic field causes the resonance frequency to decrease by $\sim$1 MHz. At a magnetic field of 9 T, the quality factor has decreased by a factor of $\sim$200, compared to zero magnetic field.

The decrease in resonance frequency and quality factor with magnetic field can be explained as follows. As the tunnel coupling is decreased through a parallel magnetic field, the charge transition takes place in a smaller range of gate voltage, yielding a larger slope $\partial\langle N\rangle/\partial V_g$. The larger change in the displacement-dependent force acting on the CNT during a mechanical oscillation causes a larger change in spring constant and therefore in resonance frequency. Also, the asymmetry of the CNT level between being above and below the Fermi level of the leads is sharper, leading to a larger retarded single-electron force acting on the CNT, and consequently a lower quality factor. 

The decrease in quality factor with magnetic field has to be put in contrast with the recently observed magnetic damping in CNT resonators\cite{Schmid2012}. Damping in Ref.~\onlinecite{Schmid2012} is measured with a CNT placed perpendicular, not parallel, to the magnetic field. There, Lorentz forces are the cause of damping, due to the perpendicular orientation of eddy currents flowing through the CNT with respect to the magnetic field. In this article, damping is increased as the magnetic field reduces the level broadening by increasing the band gap, and the modulation of the average charge due to the mechanical motion is increased.

\subsection{Mechanical detection of an excited state}
\label{subs:excitedstate}
\begin{figure}[ht]
\centering
	\includegraphics{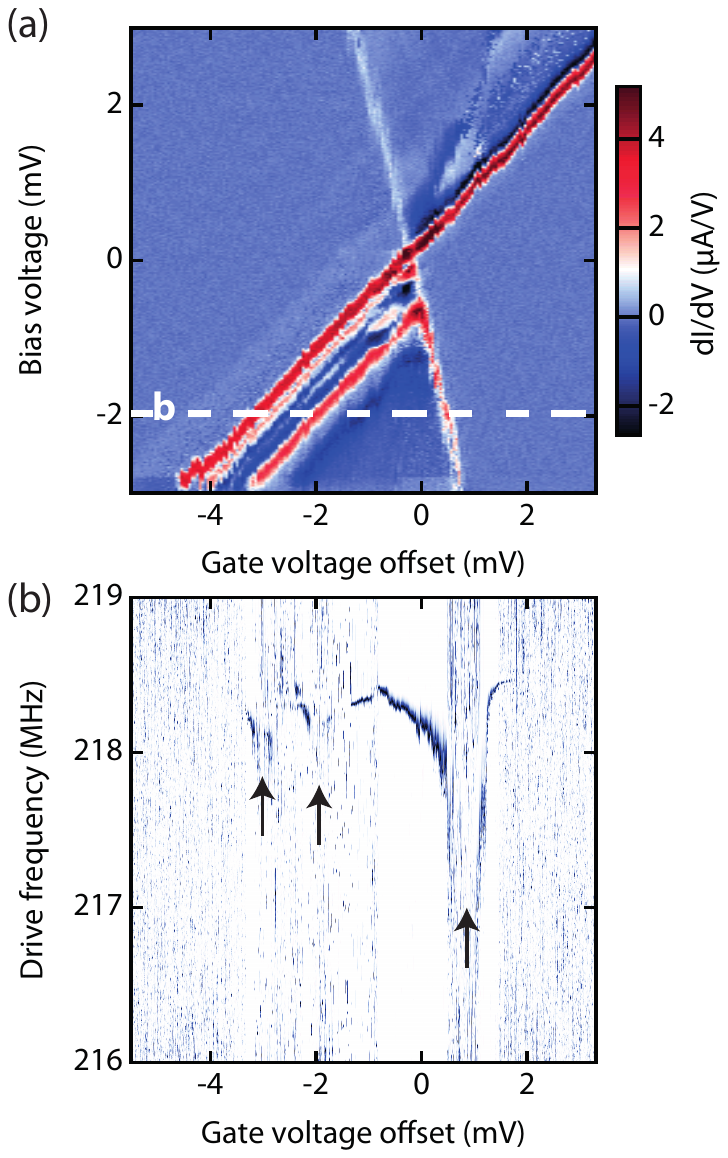}
	\caption{(a) Stability diagram at $V_g = 293$ mV of device B, showing an excited state as a diagonal line at 
negative bias voltage inside the single-electron tunneling region. (b) $|(I-I_{0})|/|I-I_0|_{max}$ as a function of drive frequency and gate voltage offset at $V_b = -2$ mV, showing three dips in resonance frequency, indicated by black arrows, with the middle dip corresponding to the excited state.	
}
\label{fig5}
\end{figure}

In this subsection, we demonstrate that the presence of excited states of the quantum dot inside the bias window can also result in frequency shifts of the mechanical resonator. Figure \ref{fig5}a shows the differential conductance, $dI/dV_b$, as a function of bias and gate voltage in the charge transition from 0 to 1 electron in a stability diagram of a different device (B). An excited electronic state of the CNT is visible as a diagonal line inside the single-electron tunneling region at negative bias. Because of charging effects, it is not possible for both states to be occupied by an electron; current still takes place through single-electron tunneling, but now can occur through two channels: through both the ground state and the excited state. This means that the rate to tunnel onto the quantum dot, $\Gamma^+$, has increased. When both states are allowed for tunneling, the current increases in a step, leading to a high $dI/dV_b$, as visible in the stability diagram.

The influence of the excited state on the mechanical motion of the CNT is examined through a measurement of the normalized current as a function of gate voltage and drive frequency at a bias voltage of -2 mV, shown in Fig. \ref{fig5}b. Three dips in resonance frequency can be seen: the leftmost and rightmost correspond to the chemical potential of the ground state aligning with the Fermi level of the source and drain, respectively. The middle dip corresponds to the chemical potential of the excited state aligning with the Fermi level of the drain. At this gate voltage, the excited state causes the rate to tunnel onto the CNT to increase, whereas the rate to tunnel off the CNT remains the same. This leads to a step in the average charge residing on the CNT. As with the ground state, tunneling through the excited state causes the CNT to experience a displacement-dependent force as it oscillates, resulting in a frequency dip. 

\subsection{Absence of single-electron spring effects in the Fabry-P\'{e}rot regime}
\label{subs:FabryPerot}
\begin{figure}[ht]
\centering
	\includegraphics{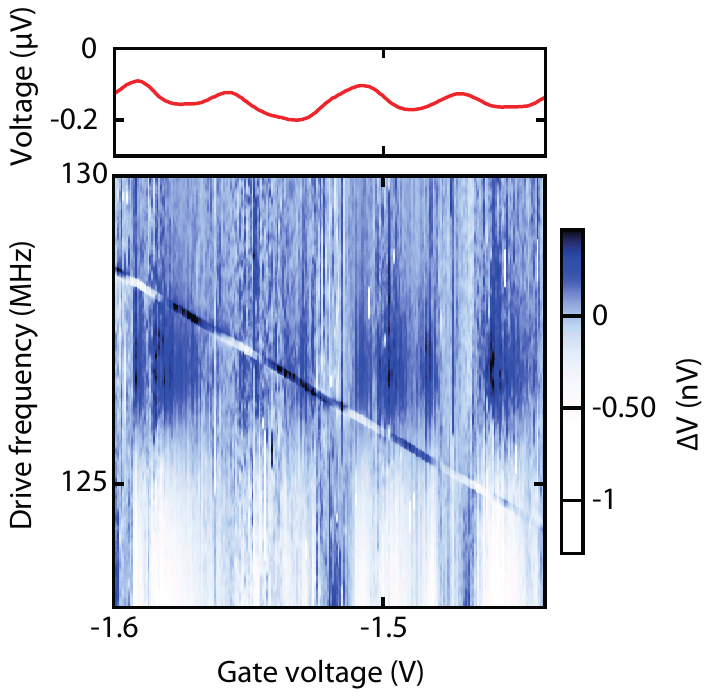}
	\caption{(top) Voltage measured across the CNT in a four terminal current bias configuration as a function of gate voltage, showing Fabry-P\'{e}rot oscillations at a bias current of -20 nA. (bottom) Change in measured voltage as a function of drive frequency and gate voltage, showing the tuning of the mechanical resonance frequency with tension arising from the gate.
}
\label{fig9}
\end{figure}

In this subsection, we examine the regime of low tunnel resistance $R_T$ for its influence on the mechanical motion by looking at the Fabry-P\'{e}rot regime\cite{Liang2001}, for which $R_T < e^2/h$. For quantum dots with large tunnel barriers ($R_T > e^2/h$), the charge on the quantum dot is quantized. For sufficiently transparent barriers with $R_T < e^2/h$, however, Coulomb blockade is destroyed by quantum fluctuations of the charge, and charge on the quantum dot is no longer quantized. In this regime, the conductance as a function of gate voltage still oscillates due to electronic Fabry-P\'{e}rot interferences, but the charge quantization is lost. Figure \ref{fig9}a shows Fabry-P\'{e}rot oscillations in the measured voltage across the CNT as a function of gate voltage using a four terminal current bias measurement of device B. Figure \ref{fig9}b displays the measured mechanical resonance frequency as a function of gate voltage and drive frequency. Visible in the plot is an increase in the resonance frequency as gate voltage is decreased, resulting from the electrostatic force from the gate that induces tension in the CNT. However, frequency dips do not occur at a Fabry-P\'{e}rot oscillation. This is because the Fabry-P\'{e}rot oscillations, in contrast to Coulomb oscillations, are not associated with discrete steps in the average charge. Softening due to the electrostatic force on the CNT quantum dot is therefore constant (or very slowly varying) across the entire gate range, and no frequency dips occur.

\section{Conclusions}
To conclude, we have found quantitative agreement between the experimental observation of single-electron effects on the resonance frequency and quality factor of a CNT quantum dot resonator, and a theoretical model. This allows the mechanical motion of a suspended CNT quantum dot to be used as a probe to detect its average charge. It is found that the mechanical resonance frequency and quality factor are reduced as the average charge changes the electrostatic force during a mechanical oscillation. At high bias, a double dip structure arises for both the resonance frequency and the quality factor, which is quantitatively supported by the model. A model, describing single-electron induced Duffing nonlinearity and mode coupling, leads to the finding of a single significant contribution. Additional experiments illustrating the model show that, by tuning the tunnel rates, the resonance frequency is reduced by $\sim$1 MHz, and the quality factor by a factor of $\sim$200. The increase of the average charge due to tunneling through an excited state of the CNT also leads to a reduction of the resonance frequency. The occurrence of frequency dips in the Fabry-P\'{e}rot regime is excluded due to the absence of steps in average charge.

\newpage
\appendix
\section{Parameters used in the model}
\label{app:param}
For the quantitative fits describing the resonance frequency, quality factor, and nonlinearity parameters in Figs. \ref{fig2}, \ref{fig:negbiasfreqfits}, \ref{fig3}, \ref{fig:comparealphas}, and \ref{fig:appothercontributions}, the following parameters were used.
\begin{table}[Ht]
	\centering
		\begin{tabular}{|c| c c|}
		\multicolumn{3}{c}{\textbf{Device properties}}\\
		\hline
		$m$&\multicolumn{2}{c|}{$2.6 \cdot 10^{-21}$ kg}\\
		 $f_0$&\multicolumn{2}{c|}{286.82 MHz}\\
		 $C_g$&\multicolumn{2}{c|}{2.9 aF}\\
		\hline
		\multicolumn{3}{c}{}\\ 
		 \multicolumn{3}{c}{\textbf{Model parameters}}\\ 
		 \hline
			figure&\ref{fig2}, \ref{fig:negbiasfreqfits}, \ref{fig:comparealphas}, and \ref{fig:appothercontributions} & \ref{fig3}\\
			\hline
			$\Gamma_{broad}$&450 GHz&540 GHz\\
		 $C_{tot}$&8.2 aF&12.3 aF\\
		 $dC_g/dx$ &-6.5 zF/nm&-5.2 zF/nm\\
		 $Q_{int}$&-&100,000\\
		 $N_0$&1&7\\
		 $B_\|$&0 T&8 T\\
		 $V_g$&0.565 V&0.9 V\\
		 $a_L$&140 GHz&22 GHz\\
		 $a_R$&170 GHz&22 GHz\\
		 $b_L$&288 /eV&20 /eV\\
		 $b_R$&288 /eV&20 /eV\\
		 \hline
		 \end{tabular}
		\caption{Properties of device A and parameters that were used for the quantitative fits in Figs. \ref{fig2} and \ref{fig3}.}
		\label{tab:param}
\end{table}
\section{Other single-electron contributions to the Duffing parameter}
\label{app:othercontributions}
\begin{figure}[Ht]
\centering
	\includegraphics{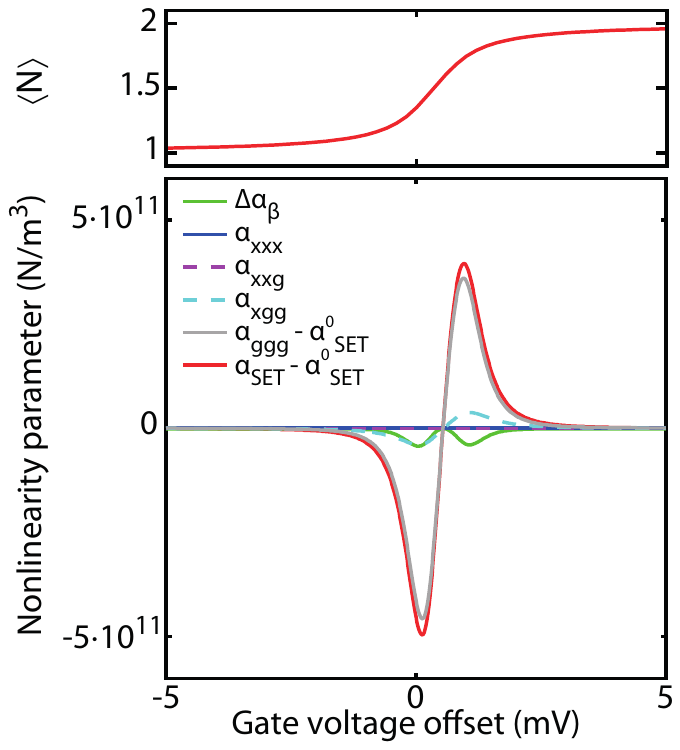}
	\caption{(top) Average occupation, $\langle N\rangle$, as a function of gate voltage, identical to the top panel of Fig. \ref{fig2}a.	(bottom) Other calculated contributions to the nonlinearity parameter, which are shown to be more than one order of magnitude smaller than $\alpha^0_{SET}$. 
}
\label{fig:appothercontributions}
\end{figure}

In subsection \ref{subs:SEDuffing}, we discussed the different single-electron contributions to the Duffing parameter, arriving at the dominance of the leading order term $\alpha^0_{SET}$. Figure \ref{fig:appothercontributions} shows that the other contributions to the single-electron Duffing parameter are more than one order of magnitude smaller and can be neglected. We use the notation where $\alpha_{ijk}$ is the contribution to $\alpha_{SET}$ resulting from the term with $F_{ijk}$ in Eq. \ref{eq:alphaandbeta}.

\newpage
\begin{acknowledgments}
We acknowledge the financial support of the Future and Emerging Technologies programme of the European Commission, under the FET-Open project QNEMS (233992), and the Foundation for Fundamental Research on Matter (FOM).
\end{acknowledgments}


\begin{thebibliography}{64}%
\makeatletter
\providecommand \@ifxundefined [1]{%
 \@ifx{#1\undefined}
}%
\providecommand \@ifnum [1]{%
 \ifnum #1\expandafter \@firstoftwo
 \else \expandafter \@secondoftwo
 \fi
}%
\providecommand \@ifx [1]{%
 \ifx #1\expandafter \@firstoftwo
 \else \expandafter \@secondoftwo
 \fi
}%
\providecommand \natexlab [1]{#1}%
\providecommand \enquote  [1]{``#1''}%
\providecommand \bibnamefont  [1]{#1}%
\providecommand \bibfnamefont [1]{#1}%
\providecommand \citenamefont [1]{#1}%
\providecommand \href@noop [0]{\@secondoftwo}%
\providecommand \href [0]{\begingroup \@sanitize@url \@href}%
\providecommand \@href[1]{\@@startlink{#1}\@@href}%
\providecommand \@@href[1]{\endgroup#1\@@endlink}%
\providecommand \@sanitize@url [0]{\catcode `\\12\catcode `\$12\catcode
  `\&12\catcode `\#12\catcode `\^12\catcode `\_12\catcode `\%12\relax}%
\providecommand \@@startlink[1]{}%
\providecommand \@@endlink[0]{}%
\providecommand \url  [0]{\begingroup\@sanitize@url \@url }%
\providecommand \@url [1]{\endgroup\@href {#1}{\urlprefix }}%
\providecommand \urlprefix  [0]{URL }%
\providecommand \Eprint [0]{\href }%
\providecommand \doibase [0]{http://dx.doi.org/}%
\providecommand \selectlanguage [0]{\@gobble}%
\providecommand \bibinfo  [0]{\@secondoftwo}%
\providecommand \bibfield  [0]{\@secondoftwo}%
\providecommand \translation [1]{[#1]}%
\providecommand \BibitemOpen [0]{}%
\providecommand \bibitemStop [0]{}%
\providecommand \bibitemNoStop [0]{.\EOS\space}%
\providecommand \EOS [0]{\spacefactor3000\relax}%
\providecommand \BibitemShut  [1]{\csname bibitem#1\endcsname}%
\let\auto@bib@innerbib\@empty
\bibitem [{\citenamefont {Craighead}(2000)}]{Craighead2000}%
  \BibitemOpen
  \bibfield  {author} {\bibinfo {author} {\bibfnamefont {H.~G.}\ \bibnamefont
  {Craighead}},\ }\href@noop {} {\bibfield  {journal} {\bibinfo  {journal}
  {Science}\ }\textbf {\bibinfo {volume} {290}},\ \bibinfo {pages} {1532}
  (\bibinfo {year} {2000})}\BibitemShut {NoStop}%
\bibitem [{\citenamefont {Ekinci}\ and\ \citenamefont
  {Roukes}(2005)}]{Ekinci2005}%
  \BibitemOpen
  \bibfield  {author} {\bibinfo {author} {\bibfnamefont {K.~L.}\ \bibnamefont
  {Ekinci}}\ and\ \bibinfo {author} {\bibfnamefont {M.~L.}\ \bibnamefont
  {Roukes}},\ }\href@noop {} {\bibfield  {journal} {\bibinfo  {journal} {Review
  Of Scientific Instruments}\ }\textbf {\bibinfo {volume} {76}},\ \bibinfo
  {pages} {061101} (\bibinfo {year} {2005})}\BibitemShut {NoStop}%
\bibitem [{\citenamefont {Lavrik}\ and\ \citenamefont
  {Datskos}(2003)}]{Lavrik2003}%
  \BibitemOpen
  \bibfield  {author} {\bibinfo {author} {\bibfnamefont {N.~V.}\ \bibnamefont
  {Lavrik}}\ and\ \bibinfo {author} {\bibfnamefont {P.~G.}\ \bibnamefont
  {Datskos}},\ }\href@noop {} {\bibfield  {journal} {\bibinfo  {journal}
  {Applied Physics Letters}\ }\textbf {\bibinfo {volume} {82}},\ \bibinfo
  {pages} {2697} (\bibinfo {year} {2003})}\BibitemShut {NoStop}%
\bibitem [{\citenamefont {Ekinci}\ \emph {et~al.}(2004)\citenamefont {Ekinci},
  \citenamefont {Huang},\ and\ \citenamefont {Roukes}}]{Ekinci2004}%
  \BibitemOpen
  \bibfield  {author} {\bibinfo {author} {\bibfnamefont {K.~L.}\ \bibnamefont
  {Ekinci}}, \bibinfo {author} {\bibfnamefont {X.~M.~H.}\ \bibnamefont
  {Huang}}, \ and\ \bibinfo {author} {\bibfnamefont {M.~L.}\ \bibnamefont
  {Roukes}},\ }\href@noop {} {\bibfield  {journal} {\bibinfo  {journal}
  {Applied Physics Letters}\ }\textbf {\bibinfo {volume} {84}},\ \bibinfo
  {pages} {4469} (\bibinfo {year} {2004})}\BibitemShut {NoStop}%
\bibitem [{\citenamefont {Lassagne}\ \emph {et~al.}(2008)\citenamefont
  {Lassagne}, \citenamefont {Garcia-Sanchez}, \citenamefont {Aguasca},\ and\
  \citenamefont {Bachtold}}]{Lassagne2008}%
  \BibitemOpen
  \bibfield  {author} {\bibinfo {author} {\bibfnamefont {B.}~\bibnamefont
  {Lassagne}}, \bibinfo {author} {\bibfnamefont {D.}~\bibnamefont
  {Garcia-Sanchez}}, \bibinfo {author} {\bibfnamefont {A.}~\bibnamefont
  {Aguasca}}, \ and\ \bibinfo {author} {\bibfnamefont {A.}~\bibnamefont
  {Bachtold}},\ }\href@noop {} {\bibfield  {journal} {\bibinfo  {journal} {Nano
  Letters}\ }\textbf {\bibinfo {volume} {8(11)}},\ \bibinfo {pages} {3735}
  (\bibinfo {year} {2008})}\BibitemShut {NoStop}%
\bibitem [{\citenamefont {Schwab}\ and\ \citenamefont
  {Roukes}(2005)}]{Schwab2005}%
  \BibitemOpen
  \bibfield  {author} {\bibinfo {author} {\bibfnamefont {K.~C.}\ \bibnamefont
  {Schwab}}\ and\ \bibinfo {author} {\bibfnamefont {M.~L.}\ \bibnamefont
  {Roukes}},\ }\href@noop {} {\bibfield  {journal} {\bibinfo  {journal}
  {Physics Today}\ }\textbf {\bibinfo {volume} {58}},\ \bibinfo {pages} {36}
  (\bibinfo {year} {2005})}\BibitemShut {NoStop}%
\bibitem [{\citenamefont {O'Connell}\ \emph {et~al.}(2010)\citenamefont
  {O'Connell}, \citenamefont {Hofheinz}, \citenamefont {Ansmann}, \citenamefont
  {Bialczak}, \citenamefont {Lenander}, \citenamefont {Lucero}, \citenamefont
  {Neeley}, \citenamefont {Sank}, \citenamefont {Wang}, \citenamefont {Weides},
  \citenamefont {Wenner}, \citenamefont {Martinis},\ and\ \citenamefont
  {Cleland}}]{O'Connell2010}%
  \BibitemOpen
  \bibfield  {author} {\bibinfo {author} {\bibfnamefont {A.~D.}\ \bibnamefont
  {O'Connell}}, \bibinfo {author} {\bibfnamefont {M.}~\bibnamefont {Hofheinz}},
  \bibinfo {author} {\bibfnamefont {M.}~\bibnamefont {Ansmann}}, \bibinfo
  {author} {\bibfnamefont {R.~C.}\ \bibnamefont {Bialczak}}, \bibinfo {author}
  {\bibfnamefont {M.}~\bibnamefont {Lenander}}, \bibinfo {author}
  {\bibfnamefont {E.}~\bibnamefont {Lucero}}, \bibinfo {author} {\bibfnamefont
  {M.}~\bibnamefont {Neeley}}, \bibinfo {author} {\bibfnamefont
  {D.}~\bibnamefont {Sank}}, \bibinfo {author} {\bibfnamefont {H.}~\bibnamefont
  {Wang}}, \bibinfo {author} {\bibfnamefont {M.}~\bibnamefont {Weides}},
  \bibinfo {author} {\bibfnamefont {J.}~\bibnamefont {Wenner}}, \bibinfo
  {author} {\bibfnamefont {J.~M.}\ \bibnamefont {Martinis}}, \ and\ \bibinfo
  {author} {\bibfnamefont {A.~N.}\ \bibnamefont {Cleland}},\ }\href@noop {}
  {\bibfield  {journal} {\bibinfo  {journal} {Nature}\ }\textbf {\bibinfo
  {volume} {464}},\ \bibinfo {pages} {697} (\bibinfo {year}
  {2010})}\BibitemShut {NoStop}%
\bibitem [{\citenamefont {Teufel}\ \emph {et~al.}(2011)\citenamefont {Teufel},
  \citenamefont {Donner}, \citenamefont {Li}, \citenamefont {Harlow},
  \citenamefont {Allman}, \citenamefont {Cicak}, \citenamefont {Sirois},
  \citenamefont {Whittaker}, \citenamefont {Lehnert},\ and\ \citenamefont
  {Simmonds}}]{Teufel2011}%
  \BibitemOpen
  \bibfield  {author} {\bibinfo {author} {\bibfnamefont {J.~D.}\ \bibnamefont
  {Teufel}}, \bibinfo {author} {\bibfnamefont {T.}~\bibnamefont {Donner}},
  \bibinfo {author} {\bibfnamefont {D.}~\bibnamefont {Li}}, \bibinfo {author}
  {\bibfnamefont {J.~W.}\ \bibnamefont {Harlow}}, \bibinfo {author}
  {\bibfnamefont {M.~S.}\ \bibnamefont {Allman}}, \bibinfo {author}
  {\bibfnamefont {K.}~\bibnamefont {Cicak}}, \bibinfo {author} {\bibfnamefont
  {A.~J.}\ \bibnamefont {Sirois}}, \bibinfo {author} {\bibfnamefont {J.~D.}\
  \bibnamefont {Whittaker}}, \bibinfo {author} {\bibfnamefont {K.~W.}\
  \bibnamefont {Lehnert}}, \ and\ \bibinfo {author} {\bibfnamefont {R.~W.}\
  \bibnamefont {Simmonds}},\ }\href@noop {} {\bibfield  {journal} {\bibinfo
  {journal} {Nature}\ }\textbf {\bibinfo {volume} {475}},\ \bibinfo {pages}
  {359} (\bibinfo {year} {2011})}\BibitemShut {NoStop}%
\bibitem [{\citenamefont {Chan}\ \emph {et~al.}(2011)\citenamefont {Chan},
  \citenamefont {Mayer~Alegre}, \citenamefont {Safavi-Naeini}, \citenamefont
  {Hill}, \citenamefont {Krause}, \citenamefont {Groeblacher}, \citenamefont
  {Aspelmeyer},\ and\ \citenamefont {Painter}}]{Chan2011}%
  \BibitemOpen
  \bibfield  {author} {\bibinfo {author} {\bibfnamefont {J.}~\bibnamefont
  {Chan}}, \bibinfo {author} {\bibfnamefont {T.~P.}\ \bibnamefont
  {Mayer~Alegre}}, \bibinfo {author} {\bibfnamefont {A.~H.}\ \bibnamefont
  {Safavi-Naeini}}, \bibinfo {author} {\bibfnamefont {J.~T.}\ \bibnamefont
  {Hill}}, \bibinfo {author} {\bibfnamefont {A.}~\bibnamefont {Krause}},
  \bibinfo {author} {\bibfnamefont {S.}~\bibnamefont {Groeblacher}}, \bibinfo
  {author} {\bibfnamefont {M.}~\bibnamefont {Aspelmeyer}}, \ and\ \bibinfo
  {author} {\bibfnamefont {O.}~\bibnamefont {Painter}},\ }\href@noop {}
  {\bibfield  {journal} {\bibinfo  {journal} {Nature}\ }\textbf {\bibinfo
  {volume} {478}},\ \bibinfo {pages} {89} (\bibinfo {year} {2011})}\BibitemShut
  {NoStop}%
\bibitem [{\citenamefont {Poot}\ and\ \citenamefont {van~der
  Zant}(2012)}]{Poot2012}%
  \BibitemOpen
  \bibfield  {author} {\bibinfo {author} {\bibfnamefont {M.}~\bibnamefont
  {Poot}}\ and\ \bibinfo {author} {\bibfnamefont {H.~S.~J.}\ \bibnamefont
  {van~der Zant}},\ }\href@noop {} {\bibfield  {journal} {\bibinfo  {journal}
  {Physics Reports-Review Section Of Physics Letters}\ }\textbf {\bibinfo
  {volume} {511}},\ \bibinfo {pages} {273} (\bibinfo {year}
  {2012})}\BibitemShut {NoStop}%
\bibitem [{\citenamefont {Knobel}\ and\ \citenamefont
  {Cleland}(2003)}]{Knobel2003}%
  \BibitemOpen
  \bibfield  {author} {\bibinfo {author} {\bibfnamefont {R.~G.}\ \bibnamefont
  {Knobel}}\ and\ \bibinfo {author} {\bibfnamefont {A.~N.}\ \bibnamefont
  {Cleland}},\ }\href@noop {} {\bibfield  {journal} {\bibinfo  {journal}
  {Nature}\ }\textbf {\bibinfo {volume} {424}},\ \bibinfo {pages} {291}
  (\bibinfo {year} {2003})}\BibitemShut {NoStop}%
\bibitem [{\citenamefont {LaHaye}\ \emph {et~al.}(2004)\citenamefont {LaHaye},
  \citenamefont {Buu}, \citenamefont {Camarota},\ and\ \citenamefont
  {Schwab}}]{LaHaye2004}%
  \BibitemOpen
  \bibfield  {author} {\bibinfo {author} {\bibfnamefont {M.~D.}\ \bibnamefont
  {LaHaye}}, \bibinfo {author} {\bibfnamefont {O.}~\bibnamefont {Buu}},
  \bibinfo {author} {\bibfnamefont {B.}~\bibnamefont {Camarota}}, \ and\
  \bibinfo {author} {\bibfnamefont {K.~C.}\ \bibnamefont {Schwab}},\
  }\href@noop {} {\bibfield  {journal} {\bibinfo  {journal} {Science}\ }\textbf
  {\bibinfo {volume} {304}},\ \bibinfo {pages} {74} (\bibinfo {year}
  {2004})}\BibitemShut {NoStop}%
\bibitem [{\citenamefont {Naik}\ \emph {et~al.}(2006)\citenamefont {Naik},
  \citenamefont {Buu}, \citenamefont {LaHaye}, \citenamefont {Armour},
  \citenamefont {Clerk}, \citenamefont {Blencowe},\ and\ \citenamefont
  {Schwab}}]{Naik2006}%
  \BibitemOpen
  \bibfield  {author} {\bibinfo {author} {\bibfnamefont {A.}~\bibnamefont
  {Naik}}, \bibinfo {author} {\bibfnamefont {O.}~\bibnamefont {Buu}}, \bibinfo
  {author} {\bibfnamefont {M.~D.}\ \bibnamefont {LaHaye}}, \bibinfo {author}
  {\bibfnamefont {A.~D.}\ \bibnamefont {Armour}}, \bibinfo {author}
  {\bibfnamefont {A.~A.}\ \bibnamefont {Clerk}}, \bibinfo {author}
  {\bibfnamefont {M.~P.}\ \bibnamefont {Blencowe}}, \ and\ \bibinfo {author}
  {\bibfnamefont {K.~C.}\ \bibnamefont {Schwab}},\ }\href@noop {} {\bibfield
  {journal} {\bibinfo  {journal} {Nature}\ }\textbf {\bibinfo {volume} {443}},\
  \bibinfo {pages} {193} (\bibinfo {year} {2006})}\BibitemShut {NoStop}%
\bibitem [{\citenamefont {Koenig}\ \emph {et~al.}(2008)\citenamefont {Koenig},
  \citenamefont {Weig},\ and\ \citenamefont {Kotthaus}}]{Koenig2008}%
  \BibitemOpen
  \bibfield  {author} {\bibinfo {author} {\bibfnamefont {D.~R.}\ \bibnamefont
  {Koenig}}, \bibinfo {author} {\bibfnamefont {E.~M.}\ \bibnamefont {Weig}}, \
  and\ \bibinfo {author} {\bibfnamefont {J.~P.}\ \bibnamefont {Kotthaus}},\
  }\href@noop {} {\bibfield  {journal} {\bibinfo  {journal} {Nature
  Nanotechnology}\ }\textbf {\bibinfo {volume} {3}},\ \bibinfo {pages} {482}
  (\bibinfo {year} {2008})}\BibitemShut {NoStop}%
\bibitem [{\citenamefont {Sazonova}\ \emph {et~al.}(2004)\citenamefont
  {Sazonova}, \citenamefont {Yaish}, \citenamefont {Ustunel}, \citenamefont
  {Roundy}, \citenamefont {Arias},\ and\ \citenamefont
  {McEuen}}]{Sazonova2004c}%
  \BibitemOpen
  \bibfield  {author} {\bibinfo {author} {\bibfnamefont {V.}~\bibnamefont
  {Sazonova}}, \bibinfo {author} {\bibfnamefont {Y.}~\bibnamefont {Yaish}},
  \bibinfo {author} {\bibfnamefont {H.}~\bibnamefont {Ustunel}}, \bibinfo
  {author} {\bibfnamefont {D.}~\bibnamefont {Roundy}}, \bibinfo {author}
  {\bibfnamefont {T.~A.}\ \bibnamefont {Arias}}, \ and\ \bibinfo {author}
  {\bibfnamefont {P.~L.}\ \bibnamefont {McEuen}},\ }\href@noop {} {\bibfield
  {journal} {\bibinfo  {journal} {Nature}\ }\textbf {\bibinfo {volume} {431}},\
  \bibinfo {pages} {284} (\bibinfo {year} {2004})}\BibitemShut {NoStop}%
\bibitem [{\citenamefont {Hüttel}\ \emph {et~al.}(2009)\citenamefont {Hüttel},
  \citenamefont {Steele}, \citenamefont {Witkamp}, \citenamefont {Poot},
  \citenamefont {Kouwenhoven},\ and\ \citenamefont {van~der
  Zant}}]{Huettel2009}%
  \BibitemOpen
  \bibfield  {author} {\bibinfo {author} {\bibfnamefont {A.~K.}\ \bibnamefont
  {Hüttel}}, \bibinfo {author} {\bibfnamefont {G.~A.}\ \bibnamefont {Steele}},
  \bibinfo {author} {\bibfnamefont {B.}~\bibnamefont {Witkamp}}, \bibinfo
  {author} {\bibfnamefont {M.}~\bibnamefont {Poot}}, \bibinfo {author}
  {\bibfnamefont {L.~P.}\ \bibnamefont {Kouwenhoven}}, \ and\ \bibinfo {author}
  {\bibfnamefont {H.~S.~J.}\ \bibnamefont {van~der Zant}},\ }\href@noop {}
  {\bibfield  {journal} {\bibinfo  {journal} {Nano Letters}\ }\textbf {\bibinfo
  {volume} {9}},\ \bibinfo {pages} {2547} (\bibinfo {year} {2009})}\BibitemShut
  {NoStop}%
\bibitem [{\citenamefont {Eichler}\ \emph {et~al.}(2011)\citenamefont
  {Eichler}, \citenamefont {Chaste}, \citenamefont {Moser},\ and\ \citenamefont
  {Bachtold}}]{Eichler2011}%
  \BibitemOpen
  \bibfield  {author} {\bibinfo {author} {\bibfnamefont {A.}~\bibnamefont
  {Eichler}}, \bibinfo {author} {\bibfnamefont {J.}~\bibnamefont {Chaste}},
  \bibinfo {author} {\bibfnamefont {J.}~\bibnamefont {Moser}}, \ and\ \bibinfo
  {author} {\bibfnamefont {A.}~\bibnamefont {Bachtold}},\ }\href {\doibase
  10.1021/nl200950d} {\bibfield  {journal} {\bibinfo  {journal} {Nano Letters}\
  } (\bibinfo {year} {2011}),\ 10.1021/nl200950d}\BibitemShut {NoStop}%
\bibitem [{\citenamefont {Schmid}\ \emph {et~al.}(2012)\citenamefont {Schmid},
  \citenamefont {Stiller}, \citenamefont {Strunk},\ and\ \citenamefont
  {Huettel}}]{Schmid2012}%
  \BibitemOpen
  \bibfield  {author} {\bibinfo {author} {\bibfnamefont {D.~R.}\ \bibnamefont
  {Schmid}}, \bibinfo {author} {\bibfnamefont {P.~L.}\ \bibnamefont {Stiller}},
  \bibinfo {author} {\bibfnamefont {C.}~\bibnamefont {Strunk}}, \ and\ \bibinfo
  {author} {\bibfnamefont {A.~K.}\ \bibnamefont {Huettel}},\ }\href@noop {}
  {\bibfield  {journal} {\bibinfo  {journal} {New Journal Of Physics}\ }\textbf
  {\bibinfo {volume} {14}},\ \bibinfo {pages} {083024} (\bibinfo {year}
  {2012})}\BibitemShut {NoStop}%
\bibitem [{\citenamefont {Tans}\ \emph {et~al.}(1997)\citenamefont {Tans},
  \citenamefont {Devoret}, \citenamefont {Dai}, \citenamefont {Thess},
  \citenamefont {Smalley}, \citenamefont {Geerligs},\ and\ \citenamefont
  {Dekker}}]{Tans1997}%
  \BibitemOpen
  \bibfield  {author} {\bibinfo {author} {\bibfnamefont {S.~J.}\ \bibnamefont
  {Tans}}, \bibinfo {author} {\bibfnamefont {M.~H.}\ \bibnamefont {Devoret}},
  \bibinfo {author} {\bibfnamefont {H.~J.}\ \bibnamefont {Dai}}, \bibinfo
  {author} {\bibfnamefont {A.}~\bibnamefont {Thess}}, \bibinfo {author}
  {\bibfnamefont {R.~E.}\ \bibnamefont {Smalley}}, \bibinfo {author}
  {\bibfnamefont {L.~J.}\ \bibnamefont {Geerligs}}, \ and\ \bibinfo {author}
  {\bibfnamefont {C.}~\bibnamefont {Dekker}},\ }\href@noop {} {\bibfield
  {journal} {\bibinfo  {journal} {Nature}\ }\textbf {\bibinfo {volume} {386}},\
  \bibinfo {pages} {474} (\bibinfo {year} {1997})}\BibitemShut {NoStop}%
\bibitem [{\citenamefont {Bockrath}\ \emph {et~al.}(1997)\citenamefont
  {Bockrath}, \citenamefont {Cobden}, \citenamefont {McEuen}, \citenamefont
  {Chopra}, \citenamefont {Zettl}, \citenamefont {Thess},\ and\ \citenamefont
  {Smalley}}]{Bockrath1997}%
  \BibitemOpen
  \bibfield  {author} {\bibinfo {author} {\bibfnamefont {M.}~\bibnamefont
  {Bockrath}}, \bibinfo {author} {\bibfnamefont {D.~H.}\ \bibnamefont
  {Cobden}}, \bibinfo {author} {\bibfnamefont {P.~L.}\ \bibnamefont {McEuen}},
  \bibinfo {author} {\bibfnamefont {N.~G.}\ \bibnamefont {Chopra}}, \bibinfo
  {author} {\bibfnamefont {A.}~\bibnamefont {Zettl}}, \bibinfo {author}
  {\bibfnamefont {A.}~\bibnamefont {Thess}}, \ and\ \bibinfo {author}
  {\bibfnamefont {R.~E.}\ \bibnamefont {Smalley}},\ }\href@noop {} {\bibfield
  {journal} {\bibinfo  {journal} {Science}\ }\textbf {\bibinfo {volume}
  {275}},\ \bibinfo {pages} {1922} (\bibinfo {year} {1997})}\BibitemShut
  {NoStop}%
\bibitem [{\citenamefont {Lassagne}\ \emph {et~al.}(2009)\citenamefont
  {Lassagne}, \citenamefont {Tarakanov}, \citenamefont {Kinaret}, \citenamefont
  {Garcia-Sanchez},\ and\ \citenamefont {Bachtold}}]{Lassagne2009}%
  \BibitemOpen
  \bibfield  {author} {\bibinfo {author} {\bibfnamefont {B.}~\bibnamefont
  {Lassagne}}, \bibinfo {author} {\bibfnamefont {Y.}~\bibnamefont {Tarakanov}},
  \bibinfo {author} {\bibfnamefont {J.}~\bibnamefont {Kinaret}}, \bibinfo
  {author} {\bibfnamefont {D.}~\bibnamefont {Garcia-Sanchez}}, \ and\ \bibinfo
  {author} {\bibfnamefont {A.}~\bibnamefont {Bachtold}},\ }\href@noop {}
  {\bibfield  {journal} {\bibinfo  {journal} {Science}\ }\textbf {\bibinfo
  {volume} {325}},\ \bibinfo {pages} {1107} (\bibinfo {year}
  {2009})}\BibitemShut {NoStop}%
\bibitem [{\citenamefont {Steele}\ \emph
  {et~al.}(2009{\natexlab{a}})\citenamefont {Steele}, \citenamefont {Hüttel},
  \citenamefont {Witkamp}, \citenamefont {Poot}, \citenamefont {Meerwaldt},
  \citenamefont {Kouwenhoven},\ and\ \citenamefont {van~der
  Zant}}]{Steele2009}%
  \BibitemOpen
  \bibfield  {author} {\bibinfo {author} {\bibfnamefont {G.~A.}\ \bibnamefont
  {Steele}}, \bibinfo {author} {\bibfnamefont {A.~K.}\ \bibnamefont {Hüttel}},
  \bibinfo {author} {\bibfnamefont {B.}~\bibnamefont {Witkamp}}, \bibinfo
  {author} {\bibfnamefont {M.}~\bibnamefont {Poot}}, \bibinfo {author}
  {\bibfnamefont {H.~B.}\ \bibnamefont {Meerwaldt}}, \bibinfo {author}
  {\bibfnamefont {L.~P.}\ \bibnamefont {Kouwenhoven}}, \ and\ \bibinfo {author}
  {\bibfnamefont {H.~S.~J.}\ \bibnamefont {van~der Zant}},\ }\href@noop {}
  {\bibfield  {journal} {\bibinfo  {journal} {Science}\ }\textbf {\bibinfo
  {volume} {325}},\ \bibinfo {pages} {1103} (\bibinfo {year}
  {2009}{\natexlab{a}})}\BibitemShut {NoStop}%
\bibitem [{\citenamefont {Meerwaldt}\ \emph {et~al.}(2012)\citenamefont
  {Meerwaldt}, \citenamefont {Steele},\ and\ \citenamefont {van~der
  Zant}}]{Meerwaldt2012}%
  \BibitemOpen
  \bibfield  {author} {\bibinfo {author} {\bibfnamefont {H.~B.}\ \bibnamefont
  {Meerwaldt}}, \bibinfo {author} {\bibfnamefont {G.~A.}\ \bibnamefont
  {Steele}}, \ and\ \bibinfo {author} {\bibfnamefont {H.~S.~J.}\ \bibnamefont
  {van~der Zant}},\ }\enquote {\bibinfo {title} {Fluctuating nonlinear
  oscillators},}\ \ (\bibinfo  {publisher} {Oxford University Press},\ \bibinfo
  {year} {2012})\ Chap.\ \bibinfo {chapter} {Carbon nanotubes: Nonlinear high-Q
  resonators with strong coupling to single-electron tunneling}, pp.\ \bibinfo
  {pages} {312--340},\ \bibinfo {note} {arXiv:1205.4921}\BibitemShut {NoStop}%
\bibitem [{\citenamefont {Doiron}\ \emph {et~al.}(2006)\citenamefont {Doiron},
  \citenamefont {Belzig},\ and\ \citenamefont {Bruder}}]{Doiron2006}%
  \BibitemOpen
  \bibfield  {author} {\bibinfo {author} {\bibfnamefont {C.~B.}\ \bibnamefont
  {Doiron}}, \bibinfo {author} {\bibfnamefont {W.}~\bibnamefont {Belzig}}, \
  and\ \bibinfo {author} {\bibfnamefont {C.}~\bibnamefont {Bruder}},\
  }\href@noop {} {\bibfield  {journal} {\bibinfo  {journal} {Physical Review
  B}\ }\textbf {\bibinfo {volume} {74}},\ \bibinfo {pages} {205336} (\bibinfo
  {year} {2006})}\BibitemShut {NoStop}%
\bibitem [{\citenamefont {Chtchelkatchev}\ \emph {et~al.}(2004)\citenamefont
  {Chtchelkatchev}, \citenamefont {Belzig},\ and\ \citenamefont
  {Bruder}}]{Chtchelkatchev2004}%
  \BibitemOpen
  \bibfield  {author} {\bibinfo {author} {\bibfnamefont {N.~M.}\ \bibnamefont
  {Chtchelkatchev}}, \bibinfo {author} {\bibfnamefont {W.}~\bibnamefont
  {Belzig}}, \ and\ \bibinfo {author} {\bibfnamefont {C.}~\bibnamefont
  {Bruder}},\ }\href@noop {} {\bibfield  {journal} {\bibinfo  {journal}
  {Physical Review B}\ }\textbf {\bibinfo {volume} {70}},\ \bibinfo {pages}
  {193305} (\bibinfo {year} {2004})}\BibitemShut {NoStop}%
\bibitem [{\citenamefont {Armour}\ \emph {et~al.}(2004)\citenamefont {Armour},
  \citenamefont {Blencowe},\ and\ \citenamefont {Zhang}}]{Armour2004}%
  \BibitemOpen
  \bibfield  {author} {\bibinfo {author} {\bibfnamefont {A.~D.}\ \bibnamefont
  {Armour}}, \bibinfo {author} {\bibfnamefont {M.~P.}\ \bibnamefont
  {Blencowe}}, \ and\ \bibinfo {author} {\bibfnamefont {Y.}~\bibnamefont
  {Zhang}},\ }\href@noop {} {\bibfield  {journal} {\bibinfo  {journal}
  {Physical Review B}\ }\textbf {\bibinfo {volume} {69}},\ \bibinfo {pages}
  {125313} (\bibinfo {year} {2004})}\BibitemShut {NoStop}%
\bibitem [{\citenamefont {Labadze}\ and\ \citenamefont
  {Blanter}(2011)}]{Labadze2011}%
  \BibitemOpen
  \bibfield  {author} {\bibinfo {author} {\bibfnamefont {G.}~\bibnamefont
  {Labadze}}\ and\ \bibinfo {author} {\bibfnamefont {Y.~M.}\ \bibnamefont
  {Blanter}},\ }\href@noop {} {\bibfield  {journal} {\bibinfo  {journal}
  {arXiv:1007.5186}\ } (\bibinfo {year} {2011})}\BibitemShut {NoStop}%
\bibitem [{\citenamefont {Armour}(2004)}]{Armour2004a}%
  \BibitemOpen
  \bibfield  {author} {\bibinfo {author} {\bibfnamefont {A.~D.}\ \bibnamefont
  {Armour}},\ }\href@noop {} {\bibfield  {journal} {\bibinfo  {journal}
  {Physical Review B}\ }\textbf {\bibinfo {volume} {70}},\ \bibinfo {pages}
  {165315} (\bibinfo {year} {2004})}\BibitemShut {NoStop}%
\bibitem [{\citenamefont {Clerk}\ and\ \citenamefont
  {Bennett}(2005)}]{Clerk2005}%
  \BibitemOpen
  \bibfield  {author} {\bibinfo {author} {\bibfnamefont {A.~A.}\ \bibnamefont
  {Clerk}}\ and\ \bibinfo {author} {\bibfnamefont {S.}~\bibnamefont
  {Bennett}},\ }\href@noop {} {\bibfield  {journal} {\bibinfo  {journal} {New
  Journal Of Physics}\ }\textbf {\bibinfo {volume} {7}},\ \bibinfo {pages}
  {238} (\bibinfo {year} {2005})}\BibitemShut {NoStop}%
\bibitem [{\citenamefont {Flindt}\ \emph {et~al.}(2005)\citenamefont {Flindt},
  \citenamefont {Novotny},\ and\ \citenamefont {Jauho}}]{Flindt2005}%
  \BibitemOpen
  \bibfield  {author} {\bibinfo {author} {\bibfnamefont {C.}~\bibnamefont
  {Flindt}}, \bibinfo {author} {\bibfnamefont {T.}~\bibnamefont {Novotny}}, \
  and\ \bibinfo {author} {\bibfnamefont {A.~P.}\ \bibnamefont {Jauho}},\
  }\href@noop {} {\bibfield  {journal} {\bibinfo  {journal} {Europhysics
  Letters}\ }\textbf {\bibinfo {volume} {69}},\ \bibinfo {pages} {475}
  (\bibinfo {year} {2005})}\BibitemShut {NoStop}%
\bibitem [{\citenamefont {Brueggemann}\ \emph {et~al.}(2012)\citenamefont
  {Brueggemann}, \citenamefont {Weick}, \citenamefont {Pistolesi},\ and\
  \citenamefont {von Oppen}}]{Brueggemann2012}%
  \BibitemOpen
  \bibfield  {author} {\bibinfo {author} {\bibfnamefont {J.}~\bibnamefont
  {Brueggemann}}, \bibinfo {author} {\bibfnamefont {G.}~\bibnamefont {Weick}},
  \bibinfo {author} {\bibfnamefont {F.}~\bibnamefont {Pistolesi}}, \ and\
  \bibinfo {author} {\bibfnamefont {F.}~\bibnamefont {von Oppen}},\ }\href@noop
  {} {\bibfield  {journal} {\bibinfo  {journal} {Physical Review B}\ }\textbf
  {\bibinfo {volume} {85}},\ \bibinfo {pages} {125441} (\bibinfo {year}
  {2012})}\BibitemShut {NoStop}%
\bibitem [{\citenamefont {Mozyrsky}\ \emph {et~al.}(2004)\citenamefont
  {Mozyrsky}, \citenamefont {Martin},\ and\ \citenamefont
  {Hastings}}]{Mozyrsky2004}%
  \BibitemOpen
  \bibfield  {author} {\bibinfo {author} {\bibfnamefont {D.}~\bibnamefont
  {Mozyrsky}}, \bibinfo {author} {\bibfnamefont {I.}~\bibnamefont {Martin}}, \
  and\ \bibinfo {author} {\bibfnamefont {M.~B.}\ \bibnamefont {Hastings}},\
  }\href@noop {} {\bibfield  {journal} {\bibinfo  {journal} {Physical Review
  Letters}\ }\textbf {\bibinfo {volume} {92}},\ \bibinfo {pages} {018303}
  (\bibinfo {year} {2004})}\BibitemShut {NoStop}%
\bibitem [{\citenamefont {Pistolesi}\ and\ \citenamefont
  {Labarthe}(2007)}]{Pistolesi2007}%
  \BibitemOpen
  \bibfield  {author} {\bibinfo {author} {\bibfnamefont {F.}~\bibnamefont
  {Pistolesi}}\ and\ \bibinfo {author} {\bibfnamefont {S.}~\bibnamefont
  {Labarthe}},\ }\href@noop {} {\bibfield  {journal} {\bibinfo  {journal}
  {Physical Review B}\ }\textbf {\bibinfo {volume} {76}},\ \bibinfo {pages}
  {165317} (\bibinfo {year} {2007})}\BibitemShut {NoStop}%
\bibitem [{\citenamefont {Pistolesi}\ \emph {et~al.}(2008)\citenamefont
  {Pistolesi}, \citenamefont {Blanter},\ and\ \citenamefont
  {Martin}}]{Pistolesi2008}%
  \BibitemOpen
  \bibfield  {author} {\bibinfo {author} {\bibfnamefont {F.}~\bibnamefont
  {Pistolesi}}, \bibinfo {author} {\bibfnamefont {Y.~M.}\ \bibnamefont
  {Blanter}}, \ and\ \bibinfo {author} {\bibfnamefont {I.}~\bibnamefont
  {Martin}},\ }\href@noop {} {\bibfield  {journal} {\bibinfo  {journal}
  {Physical Review B}\ }\textbf {\bibinfo {volume} {78}},\ \bibinfo {pages}
  {085127} (\bibinfo {year} {2008})}\BibitemShut {NoStop}%
\bibitem [{\citenamefont {Rodrigues}\ and\ \citenamefont
  {Armour}(2005)}]{Rodrigues2005}%
  \BibitemOpen
  \bibfield  {author} {\bibinfo {author} {\bibfnamefont {D.~A.}\ \bibnamefont
  {Rodrigues}}\ and\ \bibinfo {author} {\bibfnamefont {A.~D.}\ \bibnamefont
  {Armour}},\ }\href@noop {} {\bibfield  {journal} {\bibinfo  {journal} {New
  Journal Of Physics}\ }\textbf {\bibinfo {volume} {7}},\ \bibinfo {pages}
  {251} (\bibinfo {year} {2005})}\BibitemShut {NoStop}%
\bibitem [{\citenamefont {Nocera}\ \emph {et~al.}(2011)\citenamefont {Nocera},
  \citenamefont {Perroni}, \citenamefont {Marigliano~Ramaglia},\ and\
  \citenamefont {Cataudella}}]{Nocera2011}%
  \BibitemOpen
  \bibfield  {author} {\bibinfo {author} {\bibfnamefont {A.}~\bibnamefont
  {Nocera}}, \bibinfo {author} {\bibfnamefont {C.~A.}\ \bibnamefont {Perroni}},
  \bibinfo {author} {\bibfnamefont {V.}~\bibnamefont {Marigliano~Ramaglia}}, \
  and\ \bibinfo {author} {\bibfnamefont {V.}~\bibnamefont {Cataudella}},\
  }\href@noop {} {\bibfield  {journal} {\bibinfo  {journal} {arXiv:1203.2597}\
  } (\bibinfo {year} {2011})}\BibitemShut {NoStop}%
\bibitem [{\citenamefont {Rodrigues}\ \emph {et~al.}(2007)\citenamefont
  {Rodrigues}, \citenamefont {Imbers}, \citenamefont {Harvey},\ and\
  \citenamefont {Armour}}]{Rodrigues2007}%
  \BibitemOpen
  \bibfield  {author} {\bibinfo {author} {\bibfnamefont {D.~A.}\ \bibnamefont
  {Rodrigues}}, \bibinfo {author} {\bibfnamefont {J.}~\bibnamefont {Imbers}},
  \bibinfo {author} {\bibfnamefont {T.~J.}\ \bibnamefont {Harvey}}, \ and\
  \bibinfo {author} {\bibfnamefont {A.~D.}\ \bibnamefont {Armour}},\
  }\href@noop {} {\bibfield  {journal} {\bibinfo  {journal} {New Journal Of
  Physics}\ }\textbf {\bibinfo {volume} {9}},\ \bibinfo {pages} {84} (\bibinfo
  {year} {2007})}\BibitemShut {NoStop}%
\bibitem [{\citenamefont {Bennett}\ \emph {et~al.}(2010)\citenamefont
  {Bennett}, \citenamefont {Cockins}, \citenamefont {Miyahara}, \citenamefont
  {Gruetter},\ and\ \citenamefont {Clerk}}]{Bennett2010}%
  \BibitemOpen
  \bibfield  {author} {\bibinfo {author} {\bibfnamefont {S.~D.}\ \bibnamefont
  {Bennett}}, \bibinfo {author} {\bibfnamefont {L.}~\bibnamefont {Cockins}},
  \bibinfo {author} {\bibfnamefont {Y.}~\bibnamefont {Miyahara}}, \bibinfo
  {author} {\bibfnamefont {P.}~\bibnamefont {Gruetter}}, \ and\ \bibinfo
  {author} {\bibfnamefont {A.~A.}\ \bibnamefont {Clerk}},\ }\href@noop {}
  {\bibfield  {journal} {\bibinfo  {journal} {Physical Review Letters}\
  }\textbf {\bibinfo {volume} {104}},\ \bibinfo {pages} {017203} (\bibinfo
  {year} {2010})}\BibitemShut {NoStop}%
\bibitem [{\citenamefont {Ojanen}\ \emph {et~al.}(2009)\citenamefont {Ojanen},
  \citenamefont {Gethmann},\ and\ \citenamefont {von Oppen}}]{Ojanen2009}%
  \BibitemOpen
  \bibfield  {author} {\bibinfo {author} {\bibfnamefont {T.}~\bibnamefont
  {Ojanen}}, \bibinfo {author} {\bibfnamefont {F.~C.}\ \bibnamefont
  {Gethmann}}, \ and\ \bibinfo {author} {\bibfnamefont {F.}~\bibnamefont {von
  Oppen}},\ }\href@noop {} {\bibfield  {journal} {\bibinfo  {journal} {Physical
  Review B}\ }\textbf {\bibinfo {volume} {80}},\ \bibinfo {pages} {195103}
  (\bibinfo {year} {2009})}\BibitemShut {NoStop}%
\bibitem [{\citenamefont {Usmani}\ \emph {et~al.}(2007)\citenamefont {Usmani},
  \citenamefont {Blanter},\ and\ \citenamefont {Nazarov}}]{Usmani2007}%
  \BibitemOpen
  \bibfield  {author} {\bibinfo {author} {\bibfnamefont {O.}~\bibnamefont
  {Usmani}}, \bibinfo {author} {\bibfnamefont {Y.~M.}\ \bibnamefont {Blanter}},
  \ and\ \bibinfo {author} {\bibfnamefont {Y.~V.}\ \bibnamefont {Nazarov}},\
  }\href@noop {} {\bibfield  {journal} {\bibinfo  {journal} {Physical Review
  B}\ }\textbf {\bibinfo {volume} {75}},\ \bibinfo {pages} {195312} (\bibinfo
  {year} {2007})}\BibitemShut {NoStop}%
\bibitem [{\citenamefont {El~Boubsi}\ \emph {et~al.}(2008)\citenamefont
  {El~Boubsi}, \citenamefont {Usmani},\ and\ \citenamefont
  {Blanter}}]{ElBoubsi2008}%
  \BibitemOpen
  \bibfield  {author} {\bibinfo {author} {\bibfnamefont {R.}~\bibnamefont
  {El~Boubsi}}, \bibinfo {author} {\bibfnamefont {O.}~\bibnamefont {Usmani}}, \
  and\ \bibinfo {author} {\bibfnamefont {Y.~M.}\ \bibnamefont {Blanter}},\
  }\href@noop {} {\bibfield  {journal} {\bibinfo  {journal} {New Journal Of
  Physics}\ }\textbf {\bibinfo {volume} {10}},\ \bibinfo {pages} {095011}
  (\bibinfo {year} {2008})}\BibitemShut {NoStop}%
\bibitem [{\citenamefont {Midtvedt}\ \emph {et~al.}(2011)\citenamefont
  {Midtvedt}, \citenamefont {Tarakanov},\ and\ \citenamefont
  {Kinaret}}]{Midtvedt2011}%
  \BibitemOpen
  \bibfield  {author} {\bibinfo {author} {\bibfnamefont {D.}~\bibnamefont
  {Midtvedt}}, \bibinfo {author} {\bibfnamefont {Y.}~\bibnamefont {Tarakanov}},
  \ and\ \bibinfo {author} {\bibfnamefont {J.}~\bibnamefont {Kinaret}},\
  }\href@noop {} {\bibfield  {journal} {\bibinfo  {journal} {Nano Letters}\
  }\textbf {\bibinfo {volume} {11}},\ \bibinfo {pages} {1439} (\bibinfo {year}
  {2011})}\BibitemShut {NoStop}%
\bibitem [{\citenamefont {Mariani}\ and\ \citenamefont {von
  Oppen}(2009)}]{Mariani2009}%
  \BibitemOpen
  \bibfield  {author} {\bibinfo {author} {\bibfnamefont {E.}~\bibnamefont
  {Mariani}}\ and\ \bibinfo {author} {\bibfnamefont {F.}~\bibnamefont {von
  Oppen}},\ }\href@noop {} {\bibfield  {journal} {\bibinfo  {journal} {Physical
  Review B}\ }\textbf {\bibinfo {volume} {80}},\ \bibinfo {pages} {155411}
  (\bibinfo {year} {2009})}\BibitemShut {NoStop}%
\bibitem [{\citenamefont {Gorelik}\ \emph {et~al.}(1998)\citenamefont
  {Gorelik}, \citenamefont {Isacsson}, \citenamefont {Voinova}, \citenamefont
  {Kasemo}, \citenamefont {Shekhter},\ and\ \citenamefont
  {Jonson}}]{Gorelik1998}%
  \BibitemOpen
  \bibfield  {author} {\bibinfo {author} {\bibfnamefont {L.~Y.}\ \bibnamefont
  {Gorelik}}, \bibinfo {author} {\bibfnamefont {A.}~\bibnamefont {Isacsson}},
  \bibinfo {author} {\bibfnamefont {M.~V.}\ \bibnamefont {Voinova}}, \bibinfo
  {author} {\bibfnamefont {B.}~\bibnamefont {Kasemo}}, \bibinfo {author}
  {\bibfnamefont {R.~I.}\ \bibnamefont {Shekhter}}, \ and\ \bibinfo {author}
  {\bibfnamefont {M.}~\bibnamefont {Jonson}},\ }\href@noop {} {\bibfield
  {journal} {\bibinfo  {journal} {Physical Review Letters}\ }\textbf {\bibinfo
  {volume} {80}},\ \bibinfo {pages} {4526} (\bibinfo {year}
  {1998})}\BibitemShut {NoStop}%
\bibitem [{\citenamefont {Armour}\ and\ \citenamefont
  {MacKinnon}(2002)}]{Armour2002}%
  \BibitemOpen
  \bibfield  {author} {\bibinfo {author} {\bibfnamefont {A.~D.}\ \bibnamefont
  {Armour}}\ and\ \bibinfo {author} {\bibfnamefont {A.}~\bibnamefont
  {MacKinnon}},\ }\href@noop {} {\bibfield  {journal} {\bibinfo  {journal}
  {Physical Review B}\ }\textbf {\bibinfo {volume} {66}},\ \bibinfo {pages}
  {035333} (\bibinfo {year} {2002})}\BibitemShut {NoStop}%
\bibitem [{\citenamefont {Pistolesi}\ and\ \citenamefont
  {Fazio}(2005)}]{Pistolesi2005}%
  \BibitemOpen
  \bibfield  {author} {\bibinfo {author} {\bibfnamefont {F.}~\bibnamefont
  {Pistolesi}}\ and\ \bibinfo {author} {\bibfnamefont {R.}~\bibnamefont
  {Fazio}},\ }\href@noop {} {\bibfield  {journal} {\bibinfo  {journal}
  {Physical Review Letters}\ }\textbf {\bibinfo {volume} {94}},\ \bibinfo
  {pages} {036806} (\bibinfo {year} {2005})}\BibitemShut {NoStop}%
\bibitem [{\citenamefont {Huldt}\ and\ \citenamefont
  {Kinaret}(2007)}]{Huldt2007}%
  \BibitemOpen
  \bibfield  {author} {\bibinfo {author} {\bibfnamefont {C.}~\bibnamefont
  {Huldt}}\ and\ \bibinfo {author} {\bibfnamefont {J.~M.}\ \bibnamefont
  {Kinaret}},\ }\href@noop {} {\bibfield  {journal} {\bibinfo  {journal} {New
  Journal Of Physics}\ }\textbf {\bibinfo {volume} {9}},\ \bibinfo {pages} {51}
  (\bibinfo {year} {2007})}\BibitemShut {NoStop}%
\bibitem [{\citenamefont {Armour}\ \emph {et~al.}(2002)\citenamefont {Armour},
  \citenamefont {Blencowe},\ and\ \citenamefont {Schwab}}]{Armour2002a}%
  \BibitemOpen
  \bibfield  {author} {\bibinfo {author} {\bibfnamefont {A.~D.}\ \bibnamefont
  {Armour}}, \bibinfo {author} {\bibfnamefont {M.~P.}\ \bibnamefont
  {Blencowe}}, \ and\ \bibinfo {author} {\bibfnamefont {K.~C.}\ \bibnamefont
  {Schwab}},\ }\href@noop {} {\bibfield  {journal} {\bibinfo  {journal}
  {Physical Review Letters}\ }\textbf {\bibinfo {volume} {88}},\ \bibinfo
  {pages} {148301} (\bibinfo {year} {2002})}\BibitemShut {NoStop}%
\bibitem [{\citenamefont {Armour}\ and\ \citenamefont
  {Blencowe}(2008)}]{Armour2008}%
  \BibitemOpen
  \bibfield  {author} {\bibinfo {author} {\bibfnamefont {A.~D.}\ \bibnamefont
  {Armour}}\ and\ \bibinfo {author} {\bibfnamefont {M.~P.}\ \bibnamefont
  {Blencowe}},\ }\href@noop {} {\bibfield  {journal} {\bibinfo  {journal} {New
  Journal Of Physics}\ }\textbf {\bibinfo {volume} {10}},\ \bibinfo {pages}
  {095004} (\bibinfo {year} {2008})}\BibitemShut {NoStop}%
\bibitem [{\citenamefont {Blencowe}\ and\ \citenamefont
  {Armour}(2008)}]{Blencowe2008}%
  \BibitemOpen
  \bibfield  {author} {\bibinfo {author} {\bibfnamefont {M.~P.}\ \bibnamefont
  {Blencowe}}\ and\ \bibinfo {author} {\bibfnamefont {A.~D.}\ \bibnamefont
  {Armour}},\ }\href@noop {} {\bibfield  {journal} {\bibinfo  {journal} {New
  Journal Of Physics}\ }\textbf {\bibinfo {volume} {10}},\ \bibinfo {pages}
  {095005} (\bibinfo {year} {2008})}\BibitemShut {NoStop}%
\bibitem [{\citenamefont {Steele}\ \emph
  {et~al.}(2009{\natexlab{b}})\citenamefont {Steele}, \citenamefont {Gotz},\
  and\ \citenamefont {Kouwenhoven}}]{Steele2009a}%
  \BibitemOpen
  \bibfield  {author} {\bibinfo {author} {\bibfnamefont {G.~A.}\ \bibnamefont
  {Steele}}, \bibinfo {author} {\bibfnamefont {G.}~\bibnamefont {Gotz}}, \ and\
  \bibinfo {author} {\bibfnamefont {L.~P.}\ \bibnamefont {Kouwenhoven}},\
  }\href@noop {} {\bibfield  {journal} {\bibinfo  {journal} {Nature
  Nanotechnology}\ }\textbf {\bibinfo {volume} {4}},\ \bibinfo {pages} {363}
  (\bibinfo {year} {2009}{\natexlab{b}})}\BibitemShut {NoStop}%
\bibitem [{\citenamefont {Kong}\ \emph {et~al.}(1998)\citenamefont {Kong},
  \citenamefont {Soh}, \citenamefont {Cassell}, \citenamefont {Quate},\ and\
  \citenamefont {Dai}}]{Kong1998}%
  \BibitemOpen
  \bibfield  {author} {\bibinfo {author} {\bibfnamefont {J.}~\bibnamefont
  {Kong}}, \bibinfo {author} {\bibfnamefont {H.~T.}\ \bibnamefont {Soh}},
  \bibinfo {author} {\bibfnamefont {A.~M.}\ \bibnamefont {Cassell}}, \bibinfo
  {author} {\bibfnamefont {C.~F.}\ \bibnamefont {Quate}}, \ and\ \bibinfo
  {author} {\bibfnamefont {H.~J.}\ \bibnamefont {Dai}},\ }\href@noop {}
  {\bibfield  {journal} {\bibinfo  {journal} {Nature}\ }\textbf {\bibinfo
  {volume} {395}},\ \bibinfo {pages} {878} (\bibinfo {year}
  {1998})}\BibitemShut {NoStop}%
\bibitem [{\citenamefont {Liang}\ \emph {et~al.}(2001)\citenamefont {Liang},
  \citenamefont {Bockrath}, \citenamefont {Bozovic}, \citenamefont {Hafner},
  \citenamefont {Tinkham},\ and\ \citenamefont {Park}}]{Liang2001}%
  \BibitemOpen
  \bibfield  {author} {\bibinfo {author} {\bibfnamefont {W.~J.}\ \bibnamefont
  {Liang}}, \bibinfo {author} {\bibfnamefont {M.}~\bibnamefont {Bockrath}},
  \bibinfo {author} {\bibfnamefont {D.}~\bibnamefont {Bozovic}}, \bibinfo
  {author} {\bibfnamefont {J.~H.}\ \bibnamefont {Hafner}}, \bibinfo {author}
  {\bibfnamefont {M.}~\bibnamefont {Tinkham}}, \ and\ \bibinfo {author}
  {\bibfnamefont {H.}~\bibnamefont {Park}},\ }\href@noop {} {\bibfield
  {journal} {\bibinfo  {journal} {Nature}\ }\textbf {\bibinfo {volume} {411}},\
  \bibinfo {pages} {665} (\bibinfo {year} {2001})}\BibitemShut {NoStop}%
\bibitem [{\citenamefont {Beenakker}(1991)}]{Beenakker1991}%
  \BibitemOpen
  \bibfield  {author} {\bibinfo {author} {\bibfnamefont {C.~W.~J.}\
  \bibnamefont {Beenakker}},\ }\href@noop {} {\bibfield  {journal} {\bibinfo
  {journal} {Physical Review B}\ }\textbf {\bibinfo {volume} {44}},\ \bibinfo
  {pages} {1646} (\bibinfo {year} {1991})}\BibitemShut {NoStop}%
\bibitem [{\citenamefont {Thijssen}\ and\ \citenamefont {Van~der
  Zant}(2008)}]{Thijssen2008}%
  \BibitemOpen
  \bibfield  {author} {\bibinfo {author} {\bibfnamefont {J.~M.}\ \bibnamefont
  {Thijssen}}\ and\ \bibinfo {author} {\bibfnamefont {H.~S.~J.}\ \bibnamefont
  {Van~der Zant}},\ }\href@noop {} {\bibfield  {journal} {\bibinfo  {journal}
  {Physica Status Solidi B-Basic Solid State Physics}\ }\textbf {\bibinfo
  {volume} {245}},\ \bibinfo {pages} {1455} (\bibinfo {year}
  {2008})}\BibitemShut {NoStop}%
\bibitem [{\citenamefont {Korotkov}\ and\ \citenamefont
  {Nazarov}(1991)}]{Korotkov1991}%
  \BibitemOpen
  \bibfield  {author} {\bibinfo {author} {\bibfnamefont {A.~N.}\ \bibnamefont
  {Korotkov}}\ and\ \bibinfo {author} {\bibfnamefont {Y.~V.}\ \bibnamefont
  {Nazarov}},\ }\href@noop {} {\bibfield  {journal} {\bibinfo  {journal}
  {Physica B}\ }\textbf {\bibinfo {volume} {173}},\ \bibinfo {pages} {217}
  (\bibinfo {year} {1991})}\BibitemShut {NoStop}%
\bibitem [{\citenamefont {Brink}(2007)}]{Brink2007}%
  \BibitemOpen
  \bibfield  {author} {\bibinfo {author} {\bibfnamefont {M.}~\bibnamefont
  {Brink}},\ }\emph {\bibinfo {title} {Imaging single-electron charging in
  nanostructures by low-temperature scanning force microscopy}},\ \href@noop {}
  {Ph.D. thesis},\ \bibinfo  {school} {Cornell University} (\bibinfo {year}
  {2007})\BibitemShut {NoStop}%
\bibitem [{\citenamefont {Lifshitz}\ and\ \citenamefont
  {Cross}(2008)}]{Lifshitz2008}%
  \BibitemOpen
  \bibfield  {author} {\bibinfo {author} {\bibfnamefont {R.}~\bibnamefont
  {Lifshitz}}\ and\ \bibinfo {author} {\bibfnamefont {M.}~\bibnamefont
  {Cross}},\ }\enquote {\bibinfo {title} {Review of nonlinear dynamics and
  complexity},}\ \ (\bibinfo  {publisher} {John Wiley and Sons, New York,
  http://www.tau.ac.il/~ronlif/pubs/RNDC1-1-2008-preprint.pdf},\ \bibinfo
  {year} {2008})\ Chap.\ \bibinfo {chapter} {Nonlinear dynamics of
  nanomechanical and micromechanical resonators}\BibitemShut {NoStop}%
\bibitem [{\citenamefont {Castellanos-Gomez}\ \emph {et~al.}(2012)\citenamefont
  {Castellanos-Gomez}, \citenamefont {Meerwaldt}, \citenamefont {Venstra},
  \citenamefont {van~der Zant},\ and\ \citenamefont
  {Steele}}]{Castellanos-Gomez2012}%
  \BibitemOpen
  \bibfield  {author} {\bibinfo {author} {\bibfnamefont {A.}~\bibnamefont
  {Castellanos-Gomez}}, \bibinfo {author} {\bibfnamefont {H.~B.}\ \bibnamefont
  {Meerwaldt}}, \bibinfo {author} {\bibfnamefont {W.~J.}\ \bibnamefont
  {Venstra}}, \bibinfo {author} {\bibfnamefont {H.~S.~J.}\ \bibnamefont
  {van~der Zant}}, \ and\ \bibinfo {author} {\bibfnamefont {G.~A.}\
  \bibnamefont {Steele}},\ }\href {\doibase 10.1103/PhysRevB.86.041402}
  {\bibfield  {journal} {\bibinfo  {journal} {Phys. Rev. B}\ }\textbf {\bibinfo
  {volume} {86}},\ \bibinfo {pages} {041402} (\bibinfo {year}
  {2012})}\BibitemShut {NoStop}%
\bibitem [{\citenamefont {Westra}\ \emph {et~al.}(2010)\citenamefont {Westra},
  \citenamefont {Poot}, \citenamefont {van~der Zant},\ and\ \citenamefont
  {Venstra}}]{Westra2010}%
  \BibitemOpen
  \bibfield  {author} {\bibinfo {author} {\bibfnamefont {H.~J.~R.}\
  \bibnamefont {Westra}}, \bibinfo {author} {\bibfnamefont {M.}~\bibnamefont
  {Poot}}, \bibinfo {author} {\bibfnamefont {H.~S.~J.}\ \bibnamefont {van~der
  Zant}}, \ and\ \bibinfo {author} {\bibfnamefont {W.~J.}\ \bibnamefont
  {Venstra}},\ }\href@noop {} {\bibfield  {journal} {\bibinfo  {journal} {Phys
  Rev Lett}\ }\textbf {\bibinfo {volume} {105}},\ \bibinfo {pages} {117205}
  (\bibinfo {year} {2010})}\BibitemShut {NoStop}%
\bibitem [{\citenamefont {Eichler}\ \emph {et~al.}(2012)\citenamefont
  {Eichler}, \citenamefont {del~\'{A}lamo Ruiz}, \citenamefont {Plaza},\ and\
  \citenamefont {Bachtold}}]{Eichler2012}%
  \BibitemOpen
  \bibfield  {author} {\bibinfo {author} {\bibfnamefont {A.}~\bibnamefont
  {Eichler}}, \bibinfo {author} {\bibfnamefont {M.}~\bibnamefont {del~\'{A}lamo
  Ruiz}}, \bibinfo {author} {\bibfnamefont {J.}~\bibnamefont {Plaza}}, \ and\
  \bibinfo {author} {\bibfnamefont {A.}~\bibnamefont {Bachtold}},\ }\href@noop
  {} {\bibfield  {journal} {\bibinfo  {journal} {Phys Rev Lett}\ }\textbf
  {\bibinfo {volume} {109}},\ \bibinfo {pages} {025503} (\bibinfo {year}
  {2012})}\BibitemShut {NoStop}%
\bibitem [{\citenamefont {Sapmaz}\ \emph {et~al.}(2003)\citenamefont {Sapmaz},
  \citenamefont {Blanter}, \citenamefont {Gurevich},\ and\ \citenamefont
  {van~der Zant}}]{Sapmaz2003}%
  \BibitemOpen
  \bibfield  {author} {\bibinfo {author} {\bibfnamefont {S.}~\bibnamefont
  {Sapmaz}}, \bibinfo {author} {\bibfnamefont {Y.~M.}\ \bibnamefont {Blanter}},
  \bibinfo {author} {\bibfnamefont {L.}~\bibnamefont {Gurevich}}, \ and\
  \bibinfo {author} {\bibfnamefont {H.~S.~J.}\ \bibnamefont {van~der Zant}},\
  }\href@noop {} {\bibfield  {journal} {\bibinfo  {journal} {Physical Review
  B}\ }\textbf {\bibinfo {volume} {67}} (\bibinfo {year} {2003})}\BibitemShut
  {NoStop}%
\bibitem [{\citenamefont {Minot}\ \emph {et~al.}(2004)\citenamefont {Minot},
  \citenamefont {Yaish}, \citenamefont {Sazonova},\ and\ \citenamefont
  {McEuen}}]{Minot2004}%
  \BibitemOpen
  \bibfield  {author} {\bibinfo {author} {\bibfnamefont {E.~D.}\ \bibnamefont
  {Minot}}, \bibinfo {author} {\bibfnamefont {Y.}~\bibnamefont {Yaish}},
  \bibinfo {author} {\bibfnamefont {V.}~\bibnamefont {Sazonova}}, \ and\
  \bibinfo {author} {\bibfnamefont {P.~L.}\ \bibnamefont {McEuen}},\
  }\href@noop {} {\bibfield  {journal} {\bibinfo  {journal} {Nature}\ }\textbf
  {\bibinfo {volume} {428}},\ \bibinfo {pages} {536} (\bibinfo {year}
  {2004})}\BibitemShut {NoStop}%
\bibitem [{\citenamefont {Deshpande}\ \emph {et~al.}(2009)\citenamefont
  {Deshpande}, \citenamefont {Chandra}, \citenamefont {Caldwell}, \citenamefont
  {Novikov}, \citenamefont {Hone},\ and\ \citenamefont
  {Bockrath}}]{Deshpande2009}%
  \BibitemOpen
  \bibfield  {author} {\bibinfo {author} {\bibfnamefont {V.~V.}\ \bibnamefont
  {Deshpande}}, \bibinfo {author} {\bibfnamefont {B.}~\bibnamefont {Chandra}},
  \bibinfo {author} {\bibfnamefont {R.}~\bibnamefont {Caldwell}}, \bibinfo
  {author} {\bibfnamefont {D.~S.}\ \bibnamefont {Novikov}}, \bibinfo {author}
  {\bibfnamefont {J.}~\bibnamefont {Hone}}, \ and\ \bibinfo {author}
  {\bibfnamefont {M.}~\bibnamefont {Bockrath}},\ }\href@noop {} {\bibfield
  {journal} {\bibinfo  {journal} {Science}\ }\textbf {\bibinfo {volume}
  {323}},\ \bibinfo {pages} {106} (\bibinfo {year} {2009})}\BibitemShut
  {NoStop}%
\end{thebibliography}
\end{document}